\documentclass[12pt]{article}

\usepackage[margin=1.25in]{geometry}
\usepackage{setspace}
\usepackage{enumitem}
\usepackage{mathtools}
\usepackage{amsthm}
\usepackage{dsfont}
\usepackage{amssymb}
\usepackage{dsfont}
\usepackage{bm}
\usepackage{breqn}
\usepackage{color}
\usepackage[authoryear]{natbib}
\usepackage[hidelinks,breaklinks]{hyperref}

\allowdisplaybreaks

\newtheorem{lemma}{Lemma}
\newtheorem{theorem}{Theorem}
\newtheorem{assumption}{Assumption}

\newtheorem{corollary}{Corollary}
\newtheorem{definition}{Definition}

\newtheoremstyle{named}{}{}{\itshape}{}{\bfseries}{.}{.5em}{#1\thmnote{ #3}}
\theoremstyle{named}

\newtheorem*{namedassumption}{Assumption}

\newcommand{\inv}{^{-1}}

\newcommand{\M}{\mathbb{M}}
\newcommand{\N}{\mathbb{N}}
\newcommand{\R}{\mathbb{R}}

\newcommand{\argmax}{\text{argmax}}

\begin{document}

\title{A Generalized Argmax Theorem} 
\author{Gregory Fletcher Cox\footnote{Department of Economics, National University of Singapore (\href{mailto:ecsgfc@nus.edu.sg}{ecsgfc@nus.edu.sg})} 
}
\date{\today}

\maketitle

\begin{abstract}
The argmax theorem is a useful result for deriving the limiting distribution of estimators in many models. 
The conclusion of the argmax theorem states that the argmax of a sequence of stochastic processes converges in distribution to the argmax of a limiting stochastic process. 
This paper states a generalization of the argmax theorem that allows the maximization to take place over a sequence of subsets of the domain. 
If the sequence of subsets converges to a limiting subset, then the conclusion of the argmax theorem continues to hold. 
We demonstrate the usefulness of this generalization in three example models: estimating a structural break, estimating a parameter on the boundary of the parameter space, and estimating a weakly identified parameter. 
The generalized argmax theorem simplifies the proofs for existing results and can be used to prove new results in these literatures. 
\end{abstract}

\mbox{}\\
\mbox{}\\


\pagebreak
\section{Introduction}

The argmax theorem is a useful result for deriving the limiting distribution of estimators in many models. 
The argmax theorem starts with a sequence of stochastic processes, $\mathbb{M}_n(h)$, indexed by $h\in H$, that converges, in some sense, to a limiting stochastic process, $\mathbb{M}(h)$. 
The usual statement of the argmax theorem imposes conditions so that $\hat h_n\in\argmax_{h\in H} \mathbb{M}_n(h)$ converges in distribution to $\hat h\in\argmax_{h\in H}\mathbb{M}(h)$. 
The versions of the argmax theorem in \cite{Kosorok2008} and \cite{VaartWellner1996} are particularly general in that they allow $H$ to be an arbitrary metric space. 

We are concerned with generalizing the domain of maximization from $H$ to a sequence of subsets, $\Lambda_n$, converging, in some sense, to a limit subset, $\Lambda$. 
This generalization is natural because $h$ is often a local parameter that comes from rescaling the parameter space; see Section 3.1 in \cite{VaartWellner1996}. 
Corollary 5.58 in \cite{Vaart2000} states a version of the argmax theorem that covers this generalization when $H$ is finite-dimensional. 
Our first contribution is to extend this result, allowing $H$ to be an arbitrary metric space. 
Theorem \ref{LemmaNewArgmaxTheorem}, below, shows that we only need $\Lambda_n$ to converge to $\Lambda$ in the Painlev\'{e}-Kuratowski (PK) sense. 
PK convergence is a weak type of convergence defined for a sequence of subsets of a metric space. 
Assuming only a weak type of setwise convergence is important for being able to verify it in a variety of models. 

Our second contribution is to demonstrate the usefulness of the generalization in three example models. 
With finite-dimensional $H$, the value of the generalization appears to have been overlooked in these models. 
The generalized argmax theorem simplifies the proofs for existing results and can be used to prove new results in these models. 
See Section 3 for the literature related to each model. 

(1) In time series, estimating a structural break is estimating the date at which a change in the distribution of the sample occurred. 
The estimator maximizes over a discrete set of dates that are observed in the sample. 
The generalization of the argmax theorem given in Theorem \ref{LemmaNewArgmaxTheorem} is useful as the rescaled set of dates becomes denser and converges to an interval in the PK sense. 
The generalization is also useful when a trimming parameter is used to narrow the possible break dates to a middle fraction of the sample. 
Theorem \ref{LemmaNewArgmaxTheorem} implies new results when the trimming interval is misspecified to be too narrow. 

(2) When estimating a parameter that is on or near the boundary of the parameter space, the rescaled parameter space is locally approximated by a tangent cone. 
Existing derivations of the limiting distribution require an extra step showing that this approximation is negligible. 
The generalization of the argmax theorem given in Theorem \ref{LemmaNewArgmaxTheorem} eliminates this extra step, simplifying the proof. 
The generalization also covers new results in non-regular cases, when the rescaled parameter space may converge to any closed set. 

(3) When a parameter is weakly identified, the estimator converges at a slower rate than strongly identified parameters. 
To derive the limiting distribution, the rescaled parameter space needs to be scaled differently for strongly identified and weakly identified parameters. 
The generalization of the argmax theorem given in Theorem \ref{LemmaNewArgmaxTheorem} is useful for handling this scaling of the parameter space. 
The generalization also covers new results when inequalities are available that provide information about the true value of a weakly identified parameter. 

Other related papers include \cite{Ferger2004} and \cite{SeijoSen2011}, which generalize the argmax theorem to a non-unique argmax in the limit. 
More widely, there is a broadly related literature on the convergence of the value of a sequence of optimization problems following \cite{Berge1963}. 
There is also a broadly related literature on the stability of nonlinear programming problems to perturbations of the objective function and constraints following \cite{EvansGould1970}. 
\cite{KimPollard1990} appears to be the first paper to use the argmax theorem. 

Section \ref{Generalized_Argmax_Theorem_Section} states Theorem \ref{LemmaNewArgmaxTheorem}. 
Section \ref{Example_Models_Section} discusses the example models. 
Section \ref{Proof_of_Thm_1_Section} proves Theorem \ref{LemmaNewArgmaxTheorem}. 
Section \ref{Conclusion_Section} concludes. 
An \hyperref[appn]{Appendix} contains additional proofs and lemmas. 

For clarity, we briefly introduce some notation used throughout the paper. 
We write $(a,b)$ for $(a',b')'$ for stacking two column vectors, $a$ and $b$. 
Given a function, $f$, we write $f(a,b)$ for $f((a,b))$ for simplicity. 
We write $d_a$ or $d_f$ to denote the dimension of a vector, $a$, or the dimension of the range of a function, $f$. 

\section{A Generalized Argmax Theorem}\label{Generalized_Argmax_Theorem_Section}

This section states a generalization of the argmax theorem. 
We allow the maximization to take place over a sequence of subsets, $\Lambda_n$, that PK converge to a limiting set, $\Lambda$. 
We give the definition of PK convergence in a metric space $H$ equipped with metric $d$. 
For any $h\in H$ and $\Lambda\subset H$, let $d(h,\Lambda)=\inf_{g\in \Lambda}d(h,g)$. 

\begin{definition}\label{PK_convergence}
Let $\Lambda_n, \Lambda$ be subsets of a metric space, $H$, equipped with metric $d$. 
Define 
\begin{align*}
\limsup_{n\rightarrow\infty}\Lambda_n&=\{h\in H: \liminf_{n\rightarrow\infty} d(h,\Lambda_n)=0\}\\
\liminf_{n\rightarrow\infty}\Lambda_n&=\{h\in H: \lim_{n\rightarrow\infty} d(h,\Lambda_n)=0\}. 
\end{align*}
We say that $\Lambda_n$ PK converges to $\Lambda$, denoted by $\Lambda_n\rightarrow\Lambda$, if 
\[
\limsup_{n\rightarrow\infty}\Lambda_n=\liminf_{n\rightarrow\infty}\Lambda_n=\Lambda. 
\]
\end{definition}

\noindent\textbf{Remarks.} (1) Definition \ref{PK_convergence} is a standard definition given, for example, in \cite{AubinFrankowska1990}. 
It follows from the definition that PK limits are always closed. 
Note that there is no ambiguity in using the ``$\rightarrow$'' notation for PK convergence because it applies only to sequences of subsets of $H$, while convergence in $H$ applies only to elements of $H$. 
Even then, PK convergence is equivalent to convergence in $H$ when applied to sets of singletons. 

(2) PK convergence is known to be a weak type of setwise convergence. 
It is implied by other types of setwise convergence, such as Hausdorff convergence. 
In fact, PK convergence is so weak that it has a sequential compactness property: 
any sequence of subsets of a separable metric space has a PK limit along a subsequence; see Theorem 1.1.7 in \cite{AubinFrankowska1990}. 
This is a very useful property of PK convergence because it means that, in order to verify PK convergence, all we need to do is show that the limiting set does not depend on which subsequence was taken. 
\qed\medskip

We next state the generalized argmax theorem. 
Let $H$ be a metric space and let ``$\rightsquigarrow$'' denote weak convergence of random elements in $H$ using the Hoffmann-J\o{}rgensen definition of weak convergence. 
For any compact $K\subset H$, let $\ell^\infty(K)$ denote the space of real-valued bounded functions on $K$ equipped with the supremum norm. 
We also let ``$\rightsquigarrow$'' denote weak convergence in $\ell^\infty(K)$. 
Let $o_P(1)$ denote a sequence of scalar random variables that converges in outer probability to zero. 

\begin{theorem}
\label{LemmaNewArgmaxTheorem}
Let $\M_n,\M$ be stochastic processes indexed by a metric space $H$ such that $\M_n\rightsquigarrow \M$ in $\ell^\infty(K)$ for every compact $K\subset H$. 
Let $\Lambda_n,\Lambda\subset H$ be such that $\Lambda_n\rightarrow\Lambda$. 
Suppose almost all sample paths $h\mapsto\M(h)$ are continuous and possess a unique maximum over $h\in \Lambda$ at a random point $\hat h$, which as a random map in $H$ is tight. 
If the sequence $\hat h_n\in \Lambda_n$ is uniformly tight and satisfies $\M_n(\hat h_n)\ge \sup_{h\in \Lambda_n}\M_n(h)-o_P(1)$, then $\hat h_n\rightsquigarrow \hat h$ in $H$. 
\end{theorem}

\noindent\textbf{Remarks.} (1) The statement of Theorem \ref{LemmaNewArgmaxTheorem} should be compared to the statement of Theorem 3.2.2 in \cite{VaartWellner1996}. 
The primary difference is that the maximization is taken over $\Lambda_n$, which is assumed to PK converge to $\Lambda$. 
This is more general because one can take $\Lambda_n=\Lambda=H$ for all $n$ to return to the original statement. 
The trade-off for this generality is the assumption that $\mathbb{M}(h)$ is almost surely continuous. 
This seems to be a small price to pay considering the usefulness of Theorem \ref{LemmaNewArgmaxTheorem} in the example models. 
In all the example models considered in this paper, $\mathbb{M}(h)$ is continuous. 

(2) The proof of Theorem \ref{LemmaNewArgmaxTheorem} does not follow from the argument used to prove Corollary 5.58 in \cite{Vaart2000}, which assumes $H$ is finite dimensional. 
The proof of Corollary 5.58 in \cite{Vaart2000} crucially relies on the fact that closed balls are compact. 
This fact does not generalize to an arbitrary metric space. 
Section \ref{Proof_of_Thm_1_Section} gives the proof of Theorem \ref{LemmaNewArgmaxTheorem} and details the modifications required. 

(3) Lemma 9.10 in \cite{AndrewsCheng2012} is a similar generalization of the argmax theorem. 
The given proof contains a mistake. 
Specifically, \cite{AndrewsCheng2012} invoke the extended continuous mapping theorem (Theorem 1.11.1 in \cite{VaartWellner1996}) to get 
\begin{equation}
\sup_{h\in\Lambda_n\cap F}\M_n(h)\rightsquigarrow \sup_{h\in\Lambda\cap F}\M(h),  
\end{equation}
for an arbitrary closed set, $F$. 
The argument used to verify the condition of the extended continuous mapping theorem presumes $\Lambda_n\cap F\rightarrow \Lambda\cap F$, but this does not follow. 
For example, if $\Lambda_n=\{1/n, 1-1/n\}$, $\Lambda=\{0,1\}$, and $F=[0,1/2]\cup\{1\}$, then $\Lambda_n\cap F=\{1/n\}\rightarrow\{0\}\neq\{0,1\}=\Lambda\cap F$. 
Fortunately, Lemma 9.10 in \cite{AndrewsCheng2012} is still true because the proof of Theorem \ref{LemmaNewArgmaxTheorem} given in Section \ref{Proof_of_Thm_1_Section} provides a solution. 

(4) Theorem \ref{LemmaNewArgmaxTheorem} can be used to analyze the bootstrap version of an estimator. 
After conditioning on the data, Theorem \ref{LemmaNewArgmaxTheorem} can be applied almost surely to the conditional distributions. 
To be precise, this requires the conditional distribution of $\mathbb{M}_n$ to converge weakly in $\ell^\infty(K)$ to $\mathbb{M}$ almost surely. 
In this case, $\Lambda_n$ and $\Lambda$ can be random closed sets that are measurable with respect to the data and satisfy $\Lambda_n\rightarrow\Lambda$ almost surely. 
If we also assume that the conditional distribution of $\hat h_n$ is uniformly tight almost surely and that the $o_P(1)$ term conditionally converges in probability to zero almost surely, then it follows from Theorem \ref{LemmaNewArgmaxTheorem} that the conditional distribution of $\hat h_n$ converges weakly in $H$ to $\hat h$ almost surely. 
Note that standard implementations of the bootstrap are inconsistent for the types of models considered in Section \ref{Example_Models_Section} (see \cite{Andrews2000}), although modifications are possible (see \cite{CGZ2026}). 
\qed

\section{Example Models}\label{Example_Models_Section}

This section reviews three example models and shows how the generalization of Theorem \ref{LemmaNewArgmaxTheorem} is useful. 

\subsection{Structural Break Estimation}\label{structural_break_estimation_section}

Researchers using time series data often suspect the value of a parameter changed at some date in the sample. 
Structural break estimation is an attempt to estimate the date at which the change occurred. 
Structural break estimation is a commonly used statistical technique with a large statistical literature; see \cite{CasiniPerron2019} for a recent survey. 
Here, following \cite{Bai1997}, we consider a linear regression model with a break in the value of the coefficients. 
We derive the limiting distribution of the estimator of the break date using Theorem \ref{LemmaNewArgmaxTheorem}. 

We start with a sample, $(y_t,x_t)$ for $t\in\{1,...,T\}$, where $y_t$ is scalar and $x_t$ is a $p$-vector. 
The model is defined by a linear regression of $y_t$ on $x_t$ with a break in the value of the coefficients. 
Suppose 
\begin{align}
y_t&=x'_t\beta+\epsilon_t\hspace{2cm}t\in\{1,...,k_0\}\nonumber\\
y_t&=x'_t(\beta+\delta)+\epsilon_t\hspace{1.07cm}t\in\{k_0+1,...,T\}, \label{Change-Point_regression}
\end{align}
where $\beta$ and $\delta$ are unknown $p$-vectors of coefficients and $k_0$ is the unknown break date. 
For any fixed value of $k$, we can estimate $\beta$ and $\delta$ by defining $z_t=x_t\mathds{1}\{t>k\}$ and running a regression of $y_t$ on $x_t$ and $z_t$. 
Let $\hat \delta_k$ denote the estimator of $\delta$ given the value of $k$. 

For any fixed value of $k$, let $V_T(k)=\hat\delta'_k(Z'_kMZ_k)\hat\delta_k$, where $X=(x_1,...,x_T)'$, $M=I-X(X'X)^{-1}X'$, and $Z_k=(0,...,0,x_{k+1},...,x_T)'$. 
We define $\hat k$ to maximize $V_T(k)$ over a set of possible $k$ values. 
This is equivalent to minimizing the sum of squared residuals; see equation (5) in \cite{Bai1997}. 
The maximization set is given by $\Lambda_T=\{[\lambda_1 T], [\lambda_1 T]+1, ..., [\lambda_2 T]\}$, where $[\cdot]$ is the greatest integer function and where $\lambda_1\in(0,1)$ and $\lambda_2\in(\lambda_1,1)$ are trimming parameters. 
Trimming parameters are commonly used to narrow the possible values of the break date to a middle $\lambda_2-\lambda_1$ fraction of the sample. 
This can avoid some technical complications that arise near the beginning/end of the sample. 

We next discuss the break date. 
Let $v_T$ be a sequence of positive constants such that $v_T\rightarrow 0$ and $T^{1/2}v_T\rightarrow\infty$. 
We consider two possible assumptions for the timing of the break date. 
\begin{assumption}\label{change-point_date}
The break date, $k_0$, depends implicitly on $T$ and satisfies one of the following: 
\begin{enumerate}
\item[(a)] $k_0/T\rightarrow \tau$ for some $\tau\in(\lambda_1,\lambda_2)$. 
\item[(b)] $v_T^2(\lambda_2T-k_0)\rightarrow a$ for some $a\in\R$. 
\end{enumerate}
\end{assumption}
\noindent\textbf{Remarks.} (1) Assumption \ref{change-point_date}(a) requires the true break date to be strictly within the interval defined by the trimming parameters. 
In this case, the derivation of the limiting distribution of $\hat k$ is closely related to the proof of Proposition 3 in \cite{Bai1997}. 
We show how Theorem \ref{LemmaNewArgmaxTheorem} can be used to simplify the argument. 

(2) Assumption \ref{change-point_date}(b) considers the case that the true break date is near the end of the interval defined by the trimming parameters. 
When $a$ is positive, the true break date belongs to the interval but is close to the end. 
When $a$ is negative, the true break date lies outside the interval. 
One can view this as a local misspecification of Assumption \ref{change-point_date}(a) where the trimming interval has been chosen to be too narrow. 
\qed\medskip

The goal is to derive the limiting distribution of $\hat k$. 
We use $\delta_T$ to denote the true value of $\delta$ in (\ref{Change-Point_regression}). 
We consider asymptotics where $\delta_T$ depends on the sample size. 
For $i\in\{1,2\}$, let $B_i(r)$ be a $p$-dimensional multivariate Gaussian process on $[0,\infty)$ with mean zero and covariance $\mathbb{E}B_i(r)B_i(s)'=min(r,s)\Omega_i$, where $\Omega_i$ is positive definite and $B_i(r)$ is independent across $i$. 
We use the following assumptions. 
\begin{assumption}\label{change-point_conditions}
\begin{enumerate}
\item[(a)] $\delta_T=\delta_0 v_T$ for some $\delta_0\neq 0$. 
\item[(b)] $\hat k=k_0+O_P(v_T^{-2})$. 
\item[(c)] For every $J\in\{p,...,T-p\}$, $\sum_{t=1}^Jx_tx'_t$ and $\sum_{t=J+1}^Tx_tx'_t$ are positive definite with probability 1. 
\item[(d)] For every $J_T\le k_0$ such that $J_T\rightarrow\infty$, $J_T\inv\sum_{t=k_0-J_T+1}^{k_0}x_tx'_t\rightarrow Q_1$ and for every $J_T\le T-k_0$ such that $J_T\rightarrow\infty$, $J_T\inv\sum_{t=k_0+1}^{k_0+J_T} x_tx'_t\rightarrow Q_2$, where $Q_1$ and $Q_2$ are positive definite. 
\item[(e)] $k_0^{-1/2}\sum_{t=1}^{k_0}x_t\epsilon_t=O_P(1)$ and $(T-k_0)^{-1/2}\sum_{t=k_0+1}^{T}x_t\epsilon_t=O_P(1)$. 
\item[(f)] For every $C<\infty$, 
\begin{align}
v_T\sum_{t=[k_0-rv_T^{-2}]+1}^{k_0}x_t\epsilon_t&\rightsquigarrow B_1(r)\nonumber\\
v_T\sum_{t=k_0+1}^{[k_0+rv_T^{-2}]}x_t\epsilon_t&\rightsquigarrow B_2(r)\nonumber \label{change-point_second-convergence}
\end{align}
in $\ell^\infty([0,C])$ jointly, for $r\in[0,C]$. 
\end{enumerate}
\end{assumption}
\noindent\textbf{Remarks.} (1) Assumption \ref{change-point_conditions}(a) is a restriction on the break size. 
It requires the break to be small, in the sense that the break size, $\|\delta_T\|$, converges to zero. 
It also requires the break size to not be too small, in the sense that $T^{1/2}\|\delta_T\|\rightarrow\infty$. 
This assumption is commonly used in the structural break estimation literature to simplify the limiting distribution. 
For simplicity, we proceed with Assumption \ref{change-point_conditions}(a) in order to demonstrate the usefulness of Theorem \ref{LemmaNewArgmaxTheorem} in this case. 
Theorem \ref{LemmaNewArgmaxTheorem} may also be useful when the break size is fixed, as in \cite{Hinkley1970}, or when it is very small, as in \cite{ElliottMueller2007}. 

(2) Assumption \ref{change-point_conditions}(b) assumes the estimator converges at a particular rate. 
The rate depends on the break size, given by $v_T$. 
We note that the conditions in \cite{Bai1997} are sufficient for Assumption \ref{change-point_conditions}(b). 
Since our focus is on the application of the argmax theorem, we postpone the statement of this result to Lemma \ref{change_point_verify_rate} in the \hyperref[appn]{Appendix}. 

(3) Assumptions \ref{change-point_conditions}(c)-(f) are high-level conditions used to derive the limit of the localized objective function. 
Assumption \ref{change-point_conditions}(c) is a no-perfect-multicollinearity condition. 
It makes sure $\hat\delta_k$ is well defined for all $k\in\{p,...,T-p\}$. 
Assumption \ref{change-point_conditions}(d) ensures the partial sums of $x_tx'_t$ before and after the break converge to positive definite matrices. 
Assumption \ref{change-point_conditions}(e) ensures standardized sums of $x_t\epsilon_t$ have the usual rate of convergence. 
Assumption \ref{change-point_conditions}(f) ensures the partial sums of $x_t\epsilon_t$ before and after the break converge weakly to multivariate Gaussian processes. 
Note that Assumptions \ref{change-point_conditions}(d) and (f) also allow a change in the distribution of $(x_t,\epsilon_t)$ at the break date. 
It is easy to state low-level conditions for Assumptions \ref{change-point_conditions}(c)-(f) by limiting the dependence in the distribution of $(x_t,\epsilon_t)$ and requiring certain finite moments; for example see Assumptions A2-A6 in \cite{Bai1997}. 
We abstract from stating low-level conditions in order to focus on the role of the optimization set in this application of Theorem \ref{LemmaNewArgmaxTheorem}. 
\qed\medskip

Next, we localize the objective function around $k_0$ at the $v_T^{-2}$ rate. 
For every $s\in \R$, let 
\begin{equation}
\M_T(s)=V_T([k_0+sv_T^{-2}])-V_T(k_0). 
\end{equation}
Note that $\M_T(s)$ is eventually defined over every bounded set of $s$ values. 
By definition, $v_T^2(\hat k-k_0)$ maximizes $\M_T(s)$ over $s\in v_T^2(\Lambda_T-k_0)$, a discrete set of points. 
The following lemma shows that this set PK converges to $\R$ under Assumption \ref{change-point_date}(a) and PK converges to $(-\infty,a]$ under Assumption \ref{change-point_date}(b). 

\begin{lemma}\label{change-point_Kuratowski_convergence}
\begin{enumerate}
\item[(a)] Under Assumption \ref{change-point_date}(a), $v_T^2(\Lambda_T-k_0)\rightarrow \R$. 
\item[(b)] Under Assumption \ref{change-point_date}(b), $v_T^2(\Lambda_T-k_0)\rightarrow (-\infty,a]$. 
\end{enumerate}
\end{lemma}

\noindent\textbf{Remark.} Lemma \ref{change-point_Kuratowski_convergence} covers a discrete set of points becoming more dense and converging to a continuum, a common case of PK convergence. 
In this case, PK convergence follows if one can show that every neighborhood of every point in the limit contains a point from the discrete set eventually and some neighborhood of every point not in the limit does not contain a point from the discrete set eventually.
Note that the Hausdorff distance between $v_T^2(\Lambda_T-k_0)$ and the limit is infinite for both parts of Lemma \ref{change-point_Kuratowski_convergence}. Thus, Hausdorff convergence does not hold. This demonstrates the usefulness of PK convergence as a weak type of setwise convergence. 
\qed\medskip

The limit of $\M_T(s)$ is defined as 
\begin{equation}
\M(s)=\begin{cases}
-|s|\delta'_0Q_1\delta_0+2\delta'_0 B_1(|s|)&\text{ if }s\le 0\\
-s\delta'_0 Q_2\delta_0+2\delta'_0 B_2(s)&\text{ if }s>0. 
\end{cases}
\end{equation}
The following lemma gives the convergence result that we need to verify the conditions of Theorem \ref{LemmaNewArgmaxTheorem}. 
It is closely related to Lemma A.5 in \cite{Bai1997}. 

\begin{lemma}\label{Change-Point_convergence}
Under Assumption \ref{change-point_date}(a) or (b) and Assumption \ref{change-point_conditions}, for any $C>0$, 
\[
\M_T(s)\rightsquigarrow \M(s) 
\]
in $\ell^\infty([-C,C])$. 
\end{lemma}

We can now state the following result, based on Theorem \ref{LemmaNewArgmaxTheorem}. 
\begin{corollary}\label{Cor_1}
\begin{enumerate}
\item[(a)] Under Assumptions \ref{change-point_date}(a) and \ref{change-point_conditions}, 
\[
v_T^2(\hat k - k_0)=\argmax_{s\in v_T^2(\Lambda_T-k_0)}\M_T(s)\rightsquigarrow \argmax_{s\in\R} \M(s). 
\]
\item[(b)] Under Assumptions \ref{change-point_date}(b) and \ref{change-point_conditions}, 
\[
v_T^2(\hat k-k_0)=\argmax_{s\in v_T^2(\Lambda_T-k_0)}\M_T(s)\rightsquigarrow \argmax_{s\in(-\infty,a]} \M(s). 
\]
\end{enumerate}
\end{corollary}
\noindent\textbf{Remarks.} (1) Corollary \ref{Cor_1} is stated as a corollary to Theorem \ref{LemmaNewArgmaxTheorem} because the proof consists of verifying the conditions of Theorem \ref{LemmaNewArgmaxTheorem}. 
In fact, all the conditions of Theorem \ref{LemmaNewArgmaxTheorem} are easily satisfied by Lemmas \ref{change-point_Kuratowski_convergence} and \ref{Change-Point_convergence} in this section and Lemma \ref{Change-Point-uniqueness} in the \hyperref[appn]{Appendix}. 
Lemma \ref{Change-Point-uniqueness} in the \hyperref[appn]{Appendix} verifies that the argmax of the limit exists and is unique almost surely. 

(2) Part (a) gives the limiting distribution for $\hat k$ when the true break date is strictly inside the trimming interval. 
The limiting distribution coincides with the limit without any trimming, as stated in Proposition 3 in \cite{Bai1997}. 
Part (b) gives the limiting distribution for $\hat k$ when the true break date is near the right endpoint of the trimming interval. 
Instead of maximizing over all of $s\in\R$, the maximization is taken only over $s\in(-\infty,a]$. 
Thus, the presence of the trimming has an effect on the limit, by preventing the maximum from being achieved at any value larger than $a$. 

(3) The limiting distribution from part (b) is new to the literature. 
It can be used to analyze the consequences of local misspecification of the trimming interval. 
When $a\ge 0$, the trimming parameter is helpful because it prevents $\hat k$ from being too much larger than $k_0$. 
However, when $a<0$, the trimming parameter bounds the limiting distribution of $\hat k$ away from zero, introducing a significant distortion. 
This distortion clarifies the reason why $\lambda_2$ should be chosen large enough that $k_0\le [\lambda_2T]$. 

(4) The limit in part (b) with $a>0$ is similar to the limits obtained by \cite{CasiniPerron2026} and \cite{JiangWangYu2018} using in-fill asymptotics. 
These limits restrict the maximum by both an upper and a lower bound, while the limit in part (b) has only an upper bound. 
Both \cite{CasiniPerron2026} and \cite{JiangWangYu2018} argue that their limit theory provides a better approximation to the finite-sample distribution of $\hat k$, especially replicating the multi-modality of the distribution found in simulations. 
For the same reason, the limit in part (b) may improve the approximation to the finite-sample distribution over the limit in part (a). 
(The limit in part (a) is unimodal while the limit in part (b) with $a>0$ is bimodal.) 
A related limit is obtained by \cite{CasiniPerron2021} for the generalized Laplace estimator.

(5) Analogous results could be stated that allow the break date to be near the beginning of the trimming interval, or that allow only some of the regressors to have coefficients that change. 
In addition, related results could be explored in other contexts: models with two or more breaks with a minimum distance between them, as in \cite{QuPerron2007}; change-point estimation in nonparametric regression models, as in \cite{Muller1992}; or threshold regression models, as in \cite{Hansen2000} or \cite{HidalgoLeeSeo2019}. 
\qed

\subsection{Parameter on the Boundary}\label{Parameter_on_the_Boundary_Section}

Suppose we are estimating a parameter, $\theta$, that belongs to a parameter space, $\Theta$. 
Often the definition of $\Theta$ includes inequalities that provide a priori information on the possible values of $\theta$. 
That is, $\Theta=\{\theta\in\R^{d_\theta}: g(\theta)\le 0\}$ for a vector of inequalities given by $g(\theta)$. 
These inequalities can come from nonnegativity of variance parameters (\cite{Andrews1999, Andrews2002} and \cite{Ketz2018}), known signs of some coefficients in regression models (\cite{AndrewsArmstrong2017}, \cite{AndrewsGuggenberger2010}, \cite{CGZ2026}, \cite{FanShi2023}, and \cite{KetzMcCloskey2021}), or parameter restrictions in GARCH models (\cite{Andrews2001}, \cite{CNPR2022}, \cite{FrancqZakoian2007}, \cite{FrazierRenault2020}, and \cite{Pedersen2017}). 
These inequalities can also come from moment inequalities treated as overidentifying restrictions, as in \cite{MoonSchorfheide2009}. 
When $\theta$ satisfies one or more of the inequalities with equality, then this prior information is useful for estimating $\theta$ and influences the limiting distribution of the estimator. 
We consider the case that the true value of $\theta$ is on or near the boundary of $\Theta$ and use Theorem \ref{LemmaNewArgmaxTheorem} to derive the limiting distribution of the estimator. 

The parameter-on-the-boundary problem has a long history in the statistics literature; see \cite{SilvapulleSen2005} for an overview. 
Most of this literature considers the case that $\Theta$ is Chernoff regular. 
That is, $\Theta$ can be approximated by a tangent cone, $\Lambda$, near a fixed point, $\theta_0\in\Theta$; see \cite{Chernoff1954} or \cite{Geyer1994} for a formal definition. 
The literature then derives the limiting distribution of the estimator to be the projection of a normal random vector onto $\Lambda$. 
This gives rise to a mixture of normal distributions for the estimator and a chi-bar-squared distribution for the likelihood ratio statistic. 

Beginning with \cite{Feder1968} and \cite{Moran1971}, the true value of $\theta$ need not be exactly on the boundary of $\Theta$, but it may be close to the boundary. 
Let $\theta_n\in\Theta$ denote the true value of $\theta$, indexed by $n$ because it takes a sequence of values that converges to $\theta_0$ as the sample size increases. 
If $\theta_n$ is close to the boundary in the sense that $\sqrt{n}g(\theta_n)\rightarrow b$, and if $g(\theta)$ is continuously differentiable in a neighborhood of $\theta_0$ with derivative $G(\theta_0)$, then the limiting set is given by $\Lambda=\{\lambda\in\R^{d_\theta}: b+G(\theta_0)\lambda\le 0\}$. 
Note that $\Lambda$ is a polyhedral set defined by a collection of affine inequalities. 
This result is formally stated in the following lemma. 
Let $e_j$ denote the $j$th unit basis vector in $\R^{d_g}$. 

\begin{lemma}\label{Boundary-inequalities}
Suppose $g(\theta)$ is continuously differentiable in a neighborhood of $\theta_0$ with matrix of derivatives denoted by $G(\theta)$. 
Also suppose $\sqrt{n}g(\theta_n)\rightarrow b$. 
If there exists $\tilde \lambda\in\Lambda=\{\lambda\in\R^{d_\theta}: b+G(\theta_0)\lambda\le 0\}$ such that $e'_jb+e'_jG(\theta_0)\tilde \lambda<0$ for all $j\in\{1,...,d_g\}$, then 
\[
\sqrt{n}(\Theta-\theta_n)\rightarrow \Lambda. 
\]
\end{lemma}
\noindent\textbf{Remarks.} (1) Lemma \ref{Boundary-inequalities} covers a common case of PK convergence, when nonlinear constraints converge to their local linearizations. 
It is similar to Proposition 4.7.3 in \cite{SilvapulleSen2005}, except with PK convergence to a polyhedral set instead of local approximation by a cone. 
PK convergence is more general than approximation by a cone simply because it allows the limit to be any closed set. 

(2) The assumption on $\tilde\lambda\in\Lambda$ is a type of Mangasarian-Fromowitz constraint qualification. 
It ensures $\Lambda$ has a nonempty interior. 
Otherwise, when $\Lambda$ is contained within a proper linear subspace, then the limit along a subsequence may be a subset of $\Lambda$ (and which subset may depend on which subsequence was taken). 
See \cite{KaidoMolinariStoye2022} for a discussion of constraint qualifications in partially identified models. 

(3) If some of the components of $b$ are $-\infty$, then the convergence $\sqrt{n}g(\theta_n)\rightarrow b$ should be interpreted elementwise. 
The corresponding inequalities drop out from the definition of $\Lambda$ so they have no influence on the limit. \qed\medskip 

Let $\hat\theta_n$ be an estimator of $\theta$ that maximizes a random objective function, $V_n(\theta)$ over $\theta\in\Theta$. 
We next state a corollary of Theorem \ref{LemmaNewArgmaxTheorem} that gives the limiting distribution of $\hat\theta_n$ when $\theta_n$ is local to the boundary. 
We use the following high-level assumptions. 
\begin{assumption}\label{Boundary-High-Level}
\begin{enumerate}
\item[(a)] There exists a set, $\Lambda$, such that $\sqrt{n}(\Theta-\theta_n)\rightarrow \Lambda$. 
\item[(b)] There exists an almost surely continuous random objective function, $\M(h)$, such that for every compact $K\subset\R^{d_\theta}$, $V_n(\theta_n+n^{-1/2}h)\rightsquigarrow \M(h)$ in $\ell^\infty(K)$. 
\item[(c)] $\sqrt{n}(\hat\theta_n-\theta_n)=O_P(1).$ 
\item[(d)] The argmax of $\M(h)$ over $\Lambda$ is nonempty and unique almost surely. 
\end{enumerate}
\end{assumption}
\noindent\textbf{Remarks.} (1) When $\Theta$ is Chernoff regular and $\theta_n=\theta_0$, then Assumption \ref{Boundary-High-Level}(a) is satisfied with a cone $\Lambda$. 
When $\theta_n\rightarrow\theta_0$, then Assumption \ref{Boundary-High-Level}(a) can be verified using Lemma \ref{Boundary-inequalities}, above. 
Even parameter spaces that are not Chernoff regular will satisfy Assumption \ref{Boundary-High-Level}(a) for some set $\Lambda$, along a subsequence. 

(2) Assumption \ref{Boundary-High-Level}(b) requires the objective function, localized at the $n^{-1/2}$ rate, to converge to a limiting objective function. 
In regular models, the limit is quadratic in $h$. 
Assumption \ref{Boundary-High-Level}(c) requires the estimator to converge at the $n^{-1/2}$ rate. 
There are many approaches to verifying Assumptions \ref{Boundary-High-Level}(b) and (c) in various models. 
Fairly general sufficient conditions are available in \cite{VaartWellner1996}. 
These arguments are typically unaffected by $\theta_n$ being close to the boundary of $\Theta$. 
Here we abstract from stating low-level conditions in order to focus on the role of the rescaled parameter space in the argmax theorem. 

(3) Assumption \ref{Boundary-High-Level}(d) is satisfied easily when $\M(h)$ is strictly concave and $\Lambda$ is convex. 
This holds in regular cases when $\M(h)$ is quadratic with negative definite form and $\Lambda$ is a polyhedral set. 
In non-concave/convex cases, Assumption \ref{Boundary-High-Level}(d) can be verified using \cite{Cox2020}. \qed

\begin{corollary}\label{Boundary-Corollary}
Under Assumption \ref{Boundary-High-Level}, 
\[
\sqrt{n}(\hat\theta_n-\theta_n)=\argmax_{h\in\sqrt{n}(\Theta-\theta_n)}V_n(\theta_n+n^{-1/2}h)\rightsquigarrow \argmax_{h\in\Lambda}\M(h). 
\]
\end{corollary}
\noindent\textbf{Remarks.} (1) Corollary \ref{Boundary-Corollary} follows directly from Theorem \ref{LemmaNewArgmaxTheorem}. 
The focus is on the convergence of the rescaled parameter space. 
In the parameter-on-the-boundary setting, PK convergence of the rescaled parameter space follows naturally. 

(2) When $\Lambda$ is a polyhedral cone and $\M(h)$ is quadratic in $h$ with normally distributed coefficients on the linear term, then Corollary \ref{Boundary-Corollary} coincides with typical results available in the parameter-on-the-boundary literature: the limiting distribution is the projection of a normal random vector onto a cone. 
Still, the proof is arguably simpler than what is available in the literature. 
For example, Lemma 2 in \cite{Andrews1999} and Proposition 4.7.4 in \cite{SilvapulleSen2005} are challenging results to prove that are needed to show that one can replace $\sqrt{n}(\Theta-\theta_n)$ by its approximating cone. 
More generally, if $\M(h)$ is not quadratic in $h$ or $\Theta$ is not Chernoff regular, then Corollary \ref{Boundary-Corollary} is new to the literature. 

(3) The usefulness of Theorem \ref{LemmaNewArgmaxTheorem} extends beyond the case of finite-dimensional $\Theta$. 
For example, a natural estimator of a distribution function subject to shape constraints is one that minimizes a distance to the empirical distribution function. 
For the case of concave distribution functions, the minimization can be taken over the set of all concave distribution functions using an $L_p$ norm or any strictly convex norm. 
(This estimator is an alternative to the least concave majorant of the empirical distribution function.) 
The empirical distribution function is well-known to converge at the $n^{-1/2}$ rate to a Brownian bridge limit. 
If the true distribution function takes a sequence of values that converge to the boundary of the shape constraints, then Theorem \ref{LemmaNewArgmaxTheorem} implies the limiting distribution of the estimator is the argmin of the distance to a Brownian bridge over the PK limit of the rescaled shape-constrained parameter space. 
Analogous results on other shape-constrained nonparametric estimators follow from similar applications of Theorem \ref{LemmaNewArgmaxTheorem}. 
\qed

\subsection{Weak Identification}\label{Weak_Identification_Section}

A vector of parameters, $\beta$, is weakly identified if the objective function used to estimate it is flat, or nearly flat, in a neighborhood of the maximum. 
Let $V_n(\beta,\pi)$ denote a random objective function that depends on a sample of size $n$ and an additional vector of identified parameters, $\pi$. 
Suppose for some value of $\pi$, say $\pi_0$, $V_n(\beta,\pi_0)$ does not depend on the value of $\beta$. 
This designates $\pi_0$ as a problematic point in the parameter space where $\beta$ is not identified. 
Weak identification of $\beta$ arises when the true value of $\pi$, say $\pi_n$, is allowed to be a sequence that depends on the sample size and converges to $\pi_0$; see \cite{AndrewsCheng2012}, \cite{Cox_weak_id_w_bounds}, or \cite{StockWright2000}. 
Sometimes, a careful reparameterization is needed to fit a given model into this setup; see \cite{Cox2024} and \cite{HanMcCloskey2019}. 

The identified set for $\beta$ is the set of possible values that are observationally equivalent, or that cannot be distinguished from each other, based on the distribution of the sample. 
When $\pi=\pi_0$, no two values of $\beta$ can be distinguished by $V_n(\beta,\pi_0)$, and thus the identified set for $\beta$ is delineated by the boundary of the parameter space. 
Let $\Theta$ denote the parameter space for $\theta=(\beta,\pi)$. 
As in Section \ref{Parameter_on_the_Boundary_Section}, $\Theta$ is defined by a collection of inequalities that give a priori restrictions on the possible values of the parameters. 
Suppose $\Theta=\{\theta\in\R^{d_\theta}: g(\theta)\le 0\}$, for some function $g(\theta)$. 

In weak identification, these inequalities are especially important because they can provide information about the value of $\beta$. 
As a simple example, suppose $\beta$ and $\pi$ are both scalars. 
If the inequalities defined by $g(\theta)$ are $\beta\ge 0$ and $\beta\le\pi$, then the identified set for $\beta$ is given by the interval $[0,\pi]$. 
This simple example just demonstrates that the identified set for $\beta$ can be more or less informative, measured by the length of the interval, depending on the value of $\pi$. 
Most papers on weak identification exclude the possibility of informative inequalities by assuming $\Theta$ is a product space between the $\beta$ parameters and the $\pi$ parameters; see page 1060 in \cite{StockWright2000} for an example. 
Theorem \ref{LemmaNewArgmaxTheorem} can be used to derive the limiting distribution of an estimator of a weakly identified parameter without the product space assumption and thus covers weak identification with informative inequalities. 

Denote the estimators by $(\hat\beta_n,\hat\pi_n)$, which maximize $V_n(\beta,\pi)$ over $(\beta,\pi)\in\Theta$. 
The weak identification literature has derived two types of limiting distributions distinguished by the rate of convergence of $\hat\beta_n$ to the true value of $\beta$. 
Let $\beta_n$ denote the true value of $\beta$, which is allowed to be a sequence that depends on the sample size and converges to a limit, $\beta_0$. 
``Weak identification'' arises when $\hat\beta_n$ is inconsistent, while ``semi-strong identification'' arises when $\hat\beta_n$ is consistent for the true value of $\beta$ at a rate given by $a_n(\hat\beta_n-\beta_n)=O_P(1)$ for a sequence of positive constants, $a_n$, satisfying $a_n\rightarrow\infty$ and $n^{-1/2}a_n\rightarrow 0$. 

Under weak identification, the relevant scaling of the local parameter space is given by $\Lambda^{\hspace{-0.48mm}\text{W}}_n=\{(\beta,\sqrt{n}(\pi-\pi_n)): (\beta,\pi)\in\Theta\}$. 
Under semi-strong identification, the relevant scaling of the local parameter space is given by $\Lambda^{\text{SS}}_n=\{(a_n(\beta-\beta_n),\sqrt{n}(\pi-\pi_n)): (\beta,\pi)\in\Theta\}$. 
Note that for $\Lambda^{\hspace{-0.48mm}\text{W}}_n$, $\beta$ is not scaled, while for $\Lambda^{\text{SS}}_n$, $\beta$ is scaled at the $a_n$ rate, corresponding to the rate of convergence of $\hat\beta_n$. 
The following lemma gives sufficient conditions for these local parameter spaces to PK converge. 
Let $cl(A)$ and $int(A)$ denote the closure and interior of a set, $A$, respectively. 

\begin{lemma}\label{weakid-K-convergence}
\begin{enumerate}
\item[(a)] Let $\mathcal{B}^{\text{W}}=\{\beta\in\R^{d_\beta}: (\beta,\pi_0)\in\Theta\}$. 
Suppose $\Theta$ is closed and that $\mathcal{B}^{\text{W}}=cl(\{\beta\in\R^{d_\beta}: (\beta,\pi_0)\in int(\Theta)\})$. 
Then $\Lambda^{\hspace{-0.48mm}\text{W}}_n\rightarrow\mathcal{B}^{\text{W}}\times \mathbb{R}^{d_\pi}$. 

\item[(b)] Suppose $g(\beta,\pi)$ is continuously differentiable in a neighborhood of $(\beta_0,\pi_0)$ with derivative $G(\beta,\pi)=[G_\beta(\beta,\pi),G_\pi(\beta,\pi)]$. 
Also suppose $a_ng(\beta_n,\pi_n)\rightarrow b\in[-\infty,0]^{d_g}$. 
Let $\mathcal{B}^{\text{SS}}=\{\lambda\in\R^{d_\beta}: b+G_\beta(\beta_0,\pi_0)\lambda\le 0\}$ and suppose there exists a $\tilde\lambda\in\mathcal{B}^{\text{SS}}$ such that $e'_jb+e'_jG_\beta(\beta_0,\pi_0)\tilde\lambda<0$ for every $j\in\{1,...,d_g\}$. 
Then $\Lambda^{\text{SS}}_n\rightarrow \mathcal{B}^{\text{SS}}\times \R^{d_\pi}$. 
\end{enumerate}
\end{lemma}

\noindent\textbf{Remarks.} (1) Part (a) gives the limit of the rescaled parameter space under weak identification. 
Part (b) gives the limit under semi-strong identification. 
In both cases, the limits are given by Cartesian products between sets in the $\beta$ direction and $\R^{d_\pi}$ in the $\pi$ direction. 
This arises because the $\pi$ directions are scaled at a faster rate, effectively turning the inequalities so they only enforce in the $\beta$ directions. 

(2) The condition in part (a) says that $\mathcal{B}^{\text{W}}$ is the closure of the interior of $\Theta$ intersected with the set satisfying $\pi=\pi_0$. 
It mostly rules out isolated points in $\mathcal{B}^{\text{W}}$ and atypical shapes for $\Theta$. 
The condition in part (b) is a type of Mangasarian-Fromowitz constraint qualification. 

(3) The limit in part (b) is very similar to the limit of the local parameter space when the parameter is near the boundary from Lemma \ref{Boundary-inequalities} in Section \ref{Parameter_on_the_Boundary_Section}. 
Both limits are polyhedral sets, defined by a collection of affine inequalities. 
The difference is that the limit in part (b) only uses the derivative with respect to $\beta$. 
This difference comes from the different rates of convergence of $\hat\beta_n$ and $\hat\pi_n$. 
This implies a distinct change in the limit theory between strong and semi-strong identification when the parameter is near the boundary. \qed\medskip

The limiting distribution of the estimator depends on the properties of $V_n(\beta,\pi)$ in a neighborhood of the identified set. 
We use the following high-level assumptions to derive the limiting distribution using Theorem \ref{LemmaNewArgmaxTheorem}. 

\begin{namedassumption}[4(a)] Under weak identification, the following hold: 
\begin{enumerate}
\item[(i)] There exists a set, $\Lambda^{\hspace{-0.48mm}\text{W}}$, such that $\Lambda_n^{\hspace{-0.48mm}\text{W}}\rightarrow \Lambda^{\hspace{-0.48mm}\text{W}}$. 
\item[(ii)] $\sqrt{n}(\hat\pi_n-\pi_n)=O_P(1)$ and $\hat\beta_n=O_P(1)$. 
\item[(iii)] There exists an almost surely continuous random objective function, $\M^{\text{W}}(\beta,h_\pi)$ such that for all compact sets, $K$, $V_n(\beta,\pi_n+n^{-1/2}h_\pi)\rightsquigarrow \M^{\text{W}}(\beta,h_\pi)$ in $\ell^\infty(K)$ for $(\beta,h_\pi)\in K$. 
\item[(iv)] The argmax of $\M^{\text{W}}(\beta,h_\pi)$ over $(\beta,h_\pi)\in\Lambda^{\hspace{-0.48mm}\text{W}}$ is nonempty and unique almost surely. 
\end{enumerate}
\end{namedassumption}
\begin{namedassumption}[4(b)] Under semi-strong identification, the following hold: 
\begin{enumerate}
\item[(i)] There exists a set, $\Lambda^{\text{SS}}$, such that $\Lambda_n^{\text{SS}}\rightarrow \Lambda^{\text{SS}}$. 
\item[(ii)] $\sqrt{n}(\hat\pi_n-\pi_n)=O_P(1)$ and $a_n(\hat\beta_n-\beta_n)=O_P(1)$. 
\item[(iii)] There exists an almost surely continuous random objective function, $\M^{\text{SS}}(h_\beta,h_\pi)$ such that for all compact sets, $K$, $V_n(\beta_n+a_n^{-1}h_\beta,\pi_n+n^{-1/2}h_\pi)\rightsquigarrow \M^{\text{SS}}(h_\beta,h_\pi)$ in $\ell^\infty(K)$ for $(h_\beta,h_\pi)\in K$. 
\item[(iv)] The argmax of $\M^{\text{SS}}(h_\beta,h_\pi)$ over $(h_\beta,h_\pi)\in\Lambda^{\text{SS}}$ is nonempty and unique almost surely. 
\end{enumerate}
\end{namedassumption}

\noindent\textbf{Remarks.} (1) There are two different versions of Assumption 4, one for weak identification and one for semi-strong identification. 
The only difference between the two is the rescaling of the parameter space, corresponding to the rate of convergence of $\hat\beta_n$. 

(2) Part (i) of both versions assumes the rescaled parameter space PK converges. 
This holds using Lemma \ref{weakid-K-convergence} if the conditions are satisfied. 
Otherwise, it holds along a subsequence. 

(3) Parts (ii) and (iii) of both versions are high-level conditions for the rate of convergence and the limit of the standardized objective functions. 
Existing results in the weak identification literature give sufficient conditions for various objective functions. 
\cite{AndrewsCheng2014} and \cite{StockWright2000} consider a generalized method of moments objective function, \cite{AndrewsCheng2013} consider a maximum likelihood objective function, and \cite{Cox_weak_id_w_bounds} considers a minimum distance objective function. 
\cite{AndrewsCheng2012} give intermediate-level sufficient conditions for a general objective function. 
We abstract from stating the sufficient conditions here in order to focus on the role of PK convergence of the rescaled parameter space. 
For all cases considered in the above literature, $\M^{\text{SS}}(h_\beta,h_\pi)$ is quadratic in $(h_\beta,h_\pi)$ with normally distributed coefficients on the linear term. 
Also, $\M^{\text{W}}(\beta,h_\pi)$ is quadratic in $h_\pi$ but may be non-quadratic in $\beta$. 
For brevity, we abstain from stating explicit formulas for $\M^{\text{SS}}(h_\beta,h_\pi)$ and $\M^{\text{W}}(\beta,h_\pi)$. 

(4) Part (iv) for both versions is a uniqueness condition on the limit of the localized objective function maximized over the limit of the rescaled parameter space. 
When $\M^{\text{SS}}(h_\beta,h_\pi)$ is a negative definite quadratic form in $(h_\beta,h_\pi)$, this follows trivially from convexity of $\Lambda^{\text{SS}}$. 
\cite{Cox2020} gives sufficient conditions for verifying uniqueness of the argmax of $\M^{\text{W}}(\beta,h_\pi)$. \qed\medskip

The above assumptions allow us to derive the limiting distribution of the estimator using Theorem \ref{LemmaNewArgmaxTheorem}. 

\begin{corollary}\label{weakid-corollary}
\textup{(a)} Under Assumption 4(a), 
\begin{align*}
\left(\hat\beta_n,\sqrt{n}(\hat\pi_n-\pi_n)\right)&=\argmax_{(\beta,h_\pi)\in\Lambda_n^{\hspace{-0.48mm}\text{W}}}V_n(\beta,\pi_n+n^{-1/2}h_\pi)\\
&\rightsquigarrow \argmax_{(\beta,h_\pi)\in\Lambda^{\hspace{-0.48mm}\text{W}}}\M^{\text{W}}(\beta,h_\pi). 
\end{align*}
\textup{(b)} Under Assumption 4(b), 
\begin{align*}
\left(a_n(\hat\beta_n-\beta_n),\sqrt{n}(\hat\pi_n-\pi_n)\right)&=\argmax_{(h_\beta,h_\pi)\in\Lambda_n^{\text{SS}}}V_n(\beta_n+a_n^{-1}h_\beta,\pi_n+n^{-1/2}h_\pi)\\
&\rightsquigarrow \argmax_{(h_\beta,h_\pi)\in\Lambda^{\text{SS}}}\M^{\text{SS}}(h_\beta,h_\pi). 
\end{align*}
\end{corollary}

\noindent\textbf{Remark.} Corollary \ref{weakid-corollary} states an important result for the weak identification literature. 
By incorporating inequalities into the weak identification limiting distribution, hypothesis tests that are identification-robust can rely on information available in inequalities when identification is weak. 
\cite{Cox_weak_id_w_bounds} proposes such a test based on Theorem \ref{LemmaNewArgmaxTheorem} and Lemma \ref{J.12}, below. 
Other identification-robust tests available in the literature are either invalid in the presence of inequalities or do not have power against hypotheses that violate the inequalities; see the simulations in \cite{Cox_weak_id_w_bounds}. 
\qed

\section{Proof of Theorem \ref{LemmaNewArgmaxTheorem}}\label{Proof_of_Thm_1_Section}

The proof of Theorem \ref{LemmaNewArgmaxTheorem} does not follow directly from the argument used to prove Theorem 3.2.2 in \cite{VaartWellner1996}. 
A problem comes from the fact that, for an arbitrary compact set, $K$, $\Lambda_n\cap K$ need not PK converge to $\Lambda\cap K$. 
Otherwise, one could show using the extended continuous mapping theorem that $\sup_{h\in \Lambda_n\cap K}\M_n(h)\rightsquigarrow\sup_{h\in\Lambda\cap K}\M(h)$. 
The proof of Theorem \ref{LemmaNewArgmaxTheorem} would then follow from the argument used to prove Theorem 3.2.2 in \cite{VaartWellner1996}. 
The following lemma gives a solution based on bounding $K$ inside and outside by compact sets such that PK convergence holds along a subsequence. 

\begin{lemma}
\label{J.14_version_2}
Let $H$ be a metric space equipped with metric $d$. 
Suppose $\Lambda_n\rightarrow\Lambda$ for a sequence of sets $\Lambda_n$ and $\Lambda$. 
If $K$ is a compact subset of $H$, then there exist compact sets, $K_1\subset K\subset K_2\subset K_3$, and there exists a subsequence, $n_q$, such that $\Lambda_{n_q}\cap K\rightarrow \Lambda\cap K_1$ and $\Lambda_{n_q}\cap K_3\rightarrow \Lambda\cap K_2$. 
\end{lemma}

\noindent\textbf{Remark.} The proof of Lemma \ref{J.14_version_2} is simple when $H$ is finite dimensional because one can take $K_3$ to be a small closed expansion of $K$, which is compact. 
In an arbitrary metric space, $K_3$ must be defined much more carefully.
The proof is given in the \hyperref[appn]{Appendix}. 
\qed\medskip

The proof of Theorem \ref{LemmaNewArgmaxTheorem} also relies on convergence of the value function. 
The following lemma gives the statement that we need. 

\begin{lemma}
\label{J.12}
Let $\M_n,\M$ be stochastic processes indexed by a metric space $H$ such that $\M_n\rightsquigarrow \M$ in $\ell^\infty(K)$ for every compact $K\subset H$. 
Let $\Lambda_n,\Lambda\subset H$ be such that $\Lambda_n\rightarrow\Lambda$. 
Suppose that almost all sample paths $h\mapsto\M(h)$ are continuous and possess a maximum over $h\in \Lambda$ at a random point $\hat h$, which as a random map in $H$ is tight. 
If the sequence $\hat h_n\in \Lambda_n$ is uniformly tight and satisfies $\M_n(\hat h_n)\ge \sup_{h\in \Lambda_n}\M_n(h)-o_P(1)$, then $\sup_{h\in\Lambda_n}\M_n(h)\rightsquigarrow \sup_{h\in\Lambda}\M(h)$. 

Furthermore, if $\mathcal{D}$ is a metric space containing random elements $\mathbb{D}_n$ and $\mathbb{D}$, where $\mathbb{D}$ is Borel measurable with separable range, and if for every compact $K\subset H$, $\M_n\rightsquigarrow\M$ in $\ell^{\infty}(K)$ holds jointly with $\mathbb{D}_n\rightsquigarrow\mathbb{D}$, then $\sup_{h\in\Lambda_n}\M_n(h)\rightsquigarrow \sup_{h\in\Lambda}\M(h)$ holds jointly with $\mathbb{D}_n\rightsquigarrow\mathbb{D}$. 
\end{lemma}

\noindent\textbf{Remarks.} (1) The statement of Lemma \ref{J.12} is the same as Theorem \ref{LemmaNewArgmaxTheorem}, except that the conclusion is convergence of the value function and uniqueness of the maximizer in the limit is not needed. 
For completeness, the furthermore tells us that this convergence holds jointly with other weak convergence conditions. 

(2) Convergence of the value functions has independent interest. \cite{Cox_weak_id_w_bounds} uses it to derive the limit theory for likelihood ratio statistics when testing hypotheses on weakly identified parameters. 

(3) The proof of Lemma \ref{J.12} is also affected by the fact that, for an arbitrary compact set, $K$, $\Lambda_n\cap K$ need not PK converge to $\Lambda\cap K$. 
The solution is again to use Lemma \ref{J.14_version_2}, combined with a subsequencing argument. 
The proof is given in the \hyperref[appn]{Appendix}. 
\qed\medskip

We next give the proof of Theorem \ref{LemmaNewArgmaxTheorem}. 
Note how Lemmas \ref{J.14_version_2} and \ref{J.12} are used, together with a subsequencing argument. 
(The $K_2$ and $K_3$ from Lemma \ref{J.14_version_2} are not used in the proof of Theorem \ref{LemmaNewArgmaxTheorem}, but they are used in the proof of Lemma \ref{J.12}.) 
Let $P^\ast$ denote the outer probability measure. 

\begin{proof}[Proof of Theorem \ref{LemmaNewArgmaxTheorem}]
We first note that for any compact set, $K$, for any closed set $F$, and for any subsequence, $n_m$, Lemma \ref{J.14_version_2} implies that there exists a further subsequence, $n_q$, and a compact set $K_1\subset K\cap F$ such that $\Lambda_{n_q}\cap K\cap F\rightarrow \Lambda\cap K_1$. 
We can apply the extended continuous mapping theorem to show 
\begin{equation}
\sup_{h\in\Lambda_{n_q}\cap K\cap F}\M_{n_q}(h)\rightsquigarrow \sup_{h\in \Lambda\cap K_1}\M(h),\label{J.10a}
\end{equation}
where the condition of the extended continuous mapping theorem is satisfied by Lemma \ref{ECMT} in the \hyperref[appn]{Appendix}. 
We also note that, by Lemma \ref{J.12}, $\sup_{h\in\Lambda_n}\M_n(h)\rightsquigarrow \sup_{h\in\Lambda}\M(h)$, and this holds jointly with (\ref{J.10a}) along the subsequence $n_q$. 

We also note that, almost surely, 
\begin{equation}
\M(\hat h)>\sup_{h\in \Lambda\cap K\cap F} \M(h), \label{J.10b}
\end{equation}
for every compact set $K$ and for every closed set $F$ such that $\hat h\notin K\cap F$. 
If this were not true, then there would exist a sequence $\{h_m\}_{m=1}^{\infty}\subset \Lambda\cap K\cap F$ with $\M(h_m)\rightarrow \M(\hat h)$. 
Since $\Lambda\cap K\cap F$ is compact, the sequence may be chosen to be convergent; by continuity, the value $\M(h)$ at the limit would be $\M(\hat h)$. 
This contradicts the fact that $\hat h$ is unique, because $h$ is contained in the closed set $\Lambda\cap K\cap F$ and hence cannot equal $\hat h$. 

Thus, for an arbitrary closed set $F$, 
\begin{align}
&\limsup_{n\rightarrow\infty}P^*\left(\hat h_n\in F\right)\\
\le&\limsup_{n\rightarrow\infty}P^*\left(\hat h_n\in K\cap F\right)+\limsup_{n\rightarrow\infty}P^*\left(\hat h_n\notin K\right)\nonumber\\
\le&\limsup_{n\rightarrow\infty}P^*\left(\sup_{h\in \Lambda_n\cap K\cap F}\M_n(h)\ge \sup_{h\in \Lambda_n}\M_n(h)-o_P(1)\right)+\limsup_{n\rightarrow\infty}P^*\left(\hat h_n\notin K\right)\nonumber\\
\le &P\left(\sup_{h\in \Lambda\cap K\cap F}\M(h)\ge \sup_{h\in \Lambda}\M(h)\right)+\limsup_{n\rightarrow\infty}P^*\left(\hat h_n\notin K\right)\nonumber\\
\le &P\left(\biggl\{\sup_{h\in \Lambda\cap K\cap F}\M(h)\ge \sup_{h\in \Lambda\cap K}\M(h)\biggr\}\cap \{\hat h\in K\}\right)+P(\hat h\notin K)+\limsup_{n\rightarrow\infty}P^*\left(\hat h_n\notin K\right)\nonumber\\
\le &P\left(\hat h\in F\right)+P\left(\hat h\notin K\right)+\limsup_{n\rightarrow\infty}P^*\left(\hat h_n\notin K\right), \nonumber
\end{align}
where the second inequality follows from the assumed condition on $\hat h_n$, the third inequality follows from an argument in the next paragraph, and the fifth inequality follows because the event $\{\sup_{h\in \Lambda\cap K\cap F}\M(h)\ge \sup_{h\in \Lambda\cap K}\M(h)\}\cap \{\hat h\in K\}$ implies $\{\hat h\in F\}$ by (\ref{J.10b}). 
The last two terms on the right can be made arbitrarily small by the choice of $K$. 
Apply the Portmanteau Theorem to conclude that $\hat h_n\rightsquigarrow \hat h$. 

The third inequality in the above expression follows from a subsequencing argument. 
Let $n_m$ be an arbitrary subsequence. 
Then, by Lemma \ref{J.14_version_2}, there exists a further subsequence, $n_q$, and a compact set $K_1$, such that (\ref{J.10a}) holds. 
By the Portmanteau Theorem and Slutsky's Lemma, 
\begin{align}
&\limsup_{q\rightarrow\infty}P^*\left(\sup_{h\in \Lambda_{n_q}\cap K\cap F}\M_{n_q}(h)\ge \sup_{h\in \Lambda_{n_q}}\M_{n_q}(h)-o_P(1)\right)\\
&\le P\left(\sup_{h\in \Lambda\cap K_1}\M(h)\ge \sup_{h\in \Lambda}\M(h)\right)\nonumber\\
&\le P\left(\sup_{h\in \Lambda\cap K\cap F}\M(h)\ge \sup_{h\in \Lambda}\M(h)\right),\nonumber
\end{align}
where the second inequality follows from $\Lambda\cap K_1\subset \Lambda\cap K\cap F$. 
We have shown that for every subsequence, $n_m$, there exists a further subsequence, $n_q$, such that this inequality holds. 
The inequality must hold along the original sequence as well. 
\end{proof}

\section{Conclusion}\label{Conclusion_Section}

This paper states a generalization of the argmax theorem that allows the maximization to take place over a sequence of subsets of the domain. 
If the sequence of subsets PK converges to a limiting subset, then the conclusion of the argmax theorem continues to hold. 
This paper demonstrates the usefulness of this generalization in three example models: structural break estimation, estimating a parameter on the boundary, and estimating a weakly identified parameter. 
The generalized argmax theorem simplifies the proofs for existing results and can be used to prove new results in these literatures. 

\appendix

\section{Additional Proofs and Lemmas}\label{appn}

We first state some notation used in the proofs. 
In any metric space, let $B(x,\epsilon)$ denote the open ball around $x$ with radius $\epsilon$. 
For any sets, $A$ and $B$, let $\vec d(A,B)=\sup_{a\in A}d(a,B)$. 
By convention, we let $\vec d(A,B)=0$ when $A$ is empty. 

\subsection{Proofs and Lemmas for Section \ref{structural_break_estimation_section}}

\begin{proof}[Proof of Lemma \ref{change-point_Kuratowski_convergence}]
\textup{(a)} By Theorem 1.1.7 in \cite{AubinFrankowska1990} every subsequence has a further subsequence that PK converges to some limit, say $\Lambda$. 
It is sufficient to show that this $\Lambda=\R$. 
For any $x\in\R$ and for any $\epsilon>0$, if $T$ (or a subsequence in $T$) is large enough, then $v_T^2([\lambda_2T]-k_0)>x$, $v_T^2([\lambda_1T]-k_0)<x$, and $v_T^2<2\epsilon$. 
This implies that $(x-\epsilon,x+\epsilon)$ contains a point of $v_T^2(\Lambda_T-k_0)$, and therefore $x\in\Lambda$. 

\textup{(b)} The argument in part (a) applies to all $x<a$. 
For $x>a$, we simply note that, if $\epsilon\in(0,x-a)$ and if $T$ (or a subsequence in $T$) is large enough that $v_T^2([\lambda_2T]-k_0)<x-\epsilon$, then $(x-\epsilon,x+\epsilon)$ contains no elements of $v_T^2(\Lambda_T-k_0)$, and therefore $x\notin \Lambda$. 
Since $\Lambda$ is closed, $\Lambda=(-\infty,a]$. 
\end{proof}

\begin{proof}[Proof of Lemma \ref{Change-Point_convergence}]
First let $k=[k_0+sv_T^{-2}]$. 
The proof of Lemma A.5 in \cite{Bai1997} shows that several terms in an expansion of $V_T(k)-V_T(k_0)$ are asymptotically negligible. 
We show that, under Assumptions \ref{change-point_date}(a) or (b) and Assumption \ref{change-point_conditions}, each term continues to be asymptotically negligible. 
In this proof, we adopt the notation from \cite{Bai1997}, setting $R=I$, $Z_2=Z_k$, and $Z_0=Z_{k_0}$, and refer to the equations therein. 
First, consider the terms in $|k_0-k|G_T(k)$. 
We have $Z'_\Delta M Z_2=O_P(1)\|\delta_T\|^{-2}$ and $(Z'_2MZ_2)^{-1}=O_P(T^{-1})$ by Assumption \ref{change-point_conditions}(d). 
Therefore, $\delta'_T(Z'_\Delta M Z_2)(Z'_2MZ_2)^{-1}(Z'_2MZ_\Delta)\delta_T=o_P(1)$ uniformly over $s\in[-C,C]$. 
Also, by Assumption \ref{change-point_conditions}(d), 
\begin{equation}
\delta'_TZ'_\Delta X(X'X)^{-1}X'Z_\Delta\delta_T=O_P(1)/(T\|\delta_T\|^2)=o_P(1).
\end{equation}
Second, consider the terms in (A.2). 
Note that $\epsilon'MZ_0(Z'_0MZ_0)^{-1}Z'_0M\epsilon$ is $O_P(1)$ uniformly over $s\in[-C,C]$ because $\epsilon'MZ_0=O_P(T^{1/2})$ by Assumption \ref{change-point_conditions}(e) and $(Z'_0MZ_0)^{-1}=O_P(T^{-1})$ by Assumption \ref{change-point_conditions}(d). 
Also note that $\epsilon'MZ_2=\epsilon'MZ_0+O_P(\|\delta_T\|^{-1})$ by Assumption \ref{change-point_conditions}(d,e,f). 
This combines with 
\begin{equation}
(T^{-1}Z'_2MZ_2)^{-1}=(T^{-1}Z'_0MZ'_0)^{-1}+O_P(T^{-1}\|\delta_T\|^{-2})
\end{equation} 
from Assumption \ref{change-point_conditions}(d) to get that 
\begin{equation}
\epsilon'MZ_2(Z'_2MZ_2)^{-1}Z'_2M\epsilon=\epsilon'MZ_0(Z'_0MZ_0)^{-1}Z'_0M\epsilon+o_P(1). 
\end{equation}
Third, consider the terms in (A.10). 
By Assumption \ref{change-point_conditions}(d,e), 
\begin{equation}
\delta'_TZ'_\Delta M\epsilon=\delta'_TZ'_\Delta\epsilon+T^{-1/2}\|\delta_T\|^{-1}O_P(1)=\delta'_TZ'_\Delta\epsilon+o_P(1). 
\end{equation}
Also, from the calculations above, 
\begin{equation}
\delta'_T(Z'_\Delta MZ_2)(Z'_2MZ_2)^{-1}Z'_2M\epsilon=\|\delta_T\|^{-1}T^{-1/2}O_P(1)=o_P(1). 
\end{equation}
Therefore, we can write $V_T(k)-V_T(k_0)$ using the first displayed expression in the proof of Proposition 3 in \cite{Bai1997}. 
By Assumption \ref{change-point_conditions}(d,f), this converges to $\M(s)$ over $s\in[-C,C]$. 
\end{proof}

\begin{lemma}\label{change_point_verify_rate}
Suppose Assumptions A2-A6 in \cite{Bai1997} hold. 
\begin{enumerate}
\item[(a)] If Assumption \ref{change-point_date}(a) and Assumption \ref{change-point_conditions}(a) hold, then Assumption \ref{change-point_conditions}(b) holds. 
\item[(b)] If Assumption \ref{change-point_date}(b) and Assumption \ref{change-point_conditions}(a) hold, then Assumption \ref{change-point_conditions}(b) holds. 
\end{enumerate}
\end{lemma}

\begin{proof}[Proof of Lemma \ref{change_point_verify_rate}]
The proof of Lemma \ref{change_point_verify_rate} follows from modifying the proof of Proposition 1 in \cite{Bai1997}. 
Below we describe the modifications needed for parts (a) and (b). 
In this proof, we adopt the notation from \cite{Bai1997} and refer to the equations therein. 

\textup{(a)} Under Assumption \ref{change-point_date}(a), the only modification needed is to point out that $K(C)$ can be defined with $\eta=\min(\lambda_1,1-\lambda_2)>0$. 
Therefore, Lemma A.4 is not needed. 
This explains why Assumption A7 with $\alpha>0$ can be replaced by Assumption \ref{change-point_conditions}(a), which effectively sets $\alpha=0$. 
(The only place where $\alpha>0$ is used in the proof of Proposition 1 in \cite{Bai1997} is in the proof of Lemma A.4.) 

\textup{(b)} Under Assumption \ref{change-point_date}(b), we first note that, as in part (a), $K(C)$ can be defined with $\eta=\min(\lambda_1,1-\lambda_2)>0$. 
This implies that Lemma A.4 is not needed. 

Second, a problem arises that, since $k_0$ need not belong to $\Lambda_T$, we may not have $V_T(\hat k)\ge V_T(k_0)$. 
This happens when $a<0$. 
In this case, we replace $k_0$ with $\tilde k=[\lambda_2 T]$, using the fact that $\tilde k\in\Lambda_T$. 
We can then replace (A.8) with 
\begin{equation}
P\left(\sup_{|k-k_0|>C\|\delta_T\|^{-2}}V_T(k)\ge V_T(\tilde k)\right)<\epsilon. 
\end{equation}
We now use equation (A.4) to replace (A.9) with 
\begin{equation}
P\left(\sup_{k\in K(C)} \left|\frac{H_T(k)+V_T(k_0)-V_T(\tilde k)}{k_0-k}\right|>\lambda \|\delta_T\|^2\right)<\epsilon. 
\end{equation}
It is sufficient to show that 
\begin{equation}
P\left(\sup_{k\in K(C)} \left|\frac{H_T(k)}{k_0-k}\right|>\frac{\lambda}{2} \|\delta_T\|^2\right)<\epsilon/2 \label{BaiA9Part1}
\end{equation}
and 
\begin{equation}
P\left(\sup_{k\in K(C)} \left|\frac{V_T(k_0)-V_T(\tilde k)}{k_0-k}\right|>\frac{\lambda}{2} \|\delta_T\|^2\right)<\epsilon/2. \label{BaiA9Part2}
\end{equation}
(\ref{BaiA9Part1}) follows from the rest of the proof of Proposition 1 in \cite{Bai1997} with $\lambda$ replaced by $\lambda/2$ and $\epsilon$ replaced by $\epsilon/2$. 
To show (\ref{BaiA9Part2}), note that $V_T(k_0)-V_T(\tilde k)=O_P(1)$ from Lemma A.5 applied to $\tilde k$. 
($\tilde k$ belongs to $D(C)$ eventually for large enough $C$.) 
For all $k\in K(C)$, $|k-k_0|^{-1}<C^{-1}\|\delta_T\|^2$. 
Therefore, 
\begin{equation}
\sup_{k\in K(C)} \left|\frac{V_T(k_0)-V_T(\tilde k)}{k_0-k}\right|\le C^{-1}\|\delta_T\|^2O_P(1), 
\end{equation}
which implies (\ref{BaiA9Part2}) for large enough $C$. 
\end{proof}

\begin{lemma}\label{Change-Point-uniqueness}
\begin{enumerate}
\item[(a)] With probability one, there exists a unique value of $s$ that maximizes $\M(s)$ over $s\in\R$. 
\item[(b)] For all $a\in\R$, with probability one, there exists a unique value of $s$ that maximizes $\M(s)$ over $s\in(-\infty,a]$. 
\end{enumerate}
\end{lemma}

\begin{proof}[Proof of Lemma \ref{Change-Point-uniqueness}]
\textup{(a)} We first show existence. 
Note that $\M(0)=0$. 
For $s>0$, $\M(s)<0$ if $2\delta'_0B_2(s)/s<\delta'_0Q_2\delta_0$. 
Note that $\delta'_0Q_2\delta_0>0$ while by Theorem 3.30 in \cite{Klebaner2012}, $2\delta'_0B_2(s)/s\rightarrow 0$ almost surely as $s\rightarrow\infty$. 
Therefore, there exists a $C_2<\infty$ such that for all $s>C_2$, $\M(s)<0=\M(0)$. 
Similarly, for $s<0$, there exists a $C_1<\infty$ such that for all $s<-C_1$, $\M(s)<0=\M(0)$. 
Therefore, the maximization over $s\in\R$ is equal to the maximization over $s\in[-C_1,C_2]$, for which a maximizer exists because $\M(s)$ is a continuous function almost surely. 

We next show uniqueness. 
If, for every $C_1,C_2>0$, the argmax of $\M(s)$ over $s\in[-C_1,C_2]$ is unique almost surely, then the argmax of $\M(s)$ over $s\in\R$ is unique almost surely. 
This is because, if $\omega\in\Omega$ is the underlying state space, then 
\begin{align}
&\{\omega\in\Omega|\M_\omega(s)\text{ has multiple maximizers over }\R\}\\
&\hspace{1cm}=\cup_{C_1=1}^{\infty}\cup_{C_2=1}^{\infty}\{\omega\in\Omega|\M_\omega(s)\text{ has multiple maximizers over }[-C_1,C_2]\}, \nonumber
\end{align}
where $\M_\omega(s)$ denotes the draw of $\M(s)$ associated with $\omega\in\Omega$. 
Thus, we fix $C_1,C_2>0$ and show that the probability of multiple maximizers over $[-C_1,C_2]$ is zero. 

Let $Z_1=B_1(C_1)$, $Z_2=B_2(C_2)$, $G_1(r)=B_1(r)-rZ_1/C_1$, and $G_2=B_2(r)-rZ_2/C_2$, so that 
\begin{align}
B_1(r)=&G_1(r)+rZ_1/C_1\\
B_2(r)=&G_2(r)+rZ_2/C_2. \nonumber
\end{align}
Note that $Z_1$ is independent of $G_1(r)$ and $Z_2$ is independent of $G_2(r)$. 
We condition on a realization of $G_1(r)$ and $G_2(r)$ and verify the conditions of Lemma 1 in \cite{Cox2020} with respect to the randomness in $Z=(Z'_1,Z'_2)'$. 
Note that the distribution of $Z$ is normal with positive definite variance $\left[\begin{smallmatrix}C_1\Omega_1&0\\0&C_2\Omega_2\end{smallmatrix}\right]$ and, therefore, satisfies Assumption Absolute Continuity. 
Note that $[-C_1,C_2]$ is an interval and therefore satisfies Assumption Manifold. 
We can write $\M(s)$ as 
\begin{equation}
\M(s)=\begin{cases}
-|s|\delta'_0Q_1\delta_0+2\delta'_0(G_1(|s|)+|s|Z_1/C_1)&\text{ if }s\le 0\\
-s\delta'_0Q_2\delta_0+2\delta'_0(G_2(s)+sZ_2/C_2)&\text{ if }s>0, 
\end{cases}
\end{equation}
which is continuous in $s$, differentiable with respect to $Z$ with a derivative that is continuous in $s$ and $Z$. 
Therefore, Assumption Continuous Differentiability is satisfied. 
For Assumption Generic, let $s_1<s_2$. 
If $0\le s_1<s_2$, then $\frac{d}{dZ_2}(\M(s_1)-\M(s_2))=2\delta_0(s_1-s_2)/C_2\neq 0$. 
If $s_1<s_2\le 0$, then $\frac{d}{dZ_1}(\M(s_1)-\M(s_2))=2\delta_0(|s_1|-|s_2|)/C_1\neq 0$. 
If $s_1\le 0\le s_2$ with $s_1\neq 0$ or $s_2\neq 0$, then 
\begin{equation}
\frac{d}{dZ}(\M(s_1)-\M(s_2))=\left[\begin{array}{c}2\delta_0|s_1|/C_1\\-2\delta_0s_2/C_2\end{array}\right]\neq \left[\begin{array}{c}0\\0\end{array}\right]. 
\end{equation}
Therefore, Assumption Generic is satisfied. 
Therefore, by Lemma 1 in \cite{Cox2020}, $\M(s)$ has a unique maximizer over $s\in[-C_1,C_2]$ with probability 1. 

\textup{(b)} Existence and uniqueness follow from the argument in part (a) with $C_2$ replaced by $a$ when $a>0$. 
When $a\le 0$, then the existence argument continues to hold with $C_2$ replaced by $a$, and the uniqueness argument holds by taking derivatives with respect to $Z_1$ only. 
\end{proof}

\subsection{Proofs for Section \ref{Parameter_on_the_Boundary_Section}}

\begin{proof}[Proof of Lemma \ref{Boundary-inequalities}]
The condition on $\tilde\lambda$ implies the interior of $\Lambda$ is given by $\Lambda^o=\{\lambda\in\R^{d_\theta}: e'_jb+e'_jG(\theta_0)\lambda<0 \text{ for all }j=1,...,d_g\}$. 
It also implies that $\Lambda$ is equal to the closure of $\Lambda^o$. 

Let $\lambda\in\Lambda^o$ and note that, by a mean value expansion, $\tilde\theta_n=\theta_n+n^{-1/2}\lambda$ satisfies 
\begin{equation}
\sqrt{n}e'_jg(\tilde\theta_n)=\sqrt{n}e'_jg(\theta_n)+e'_jG(\ddot\theta_n)\lambda\rightarrow e'_j b+e'_j G(\theta_0)\lambda<0 \label{MVE}
\end{equation}
for some $\ddot\theta_n$ between $\tilde\theta_n$ and $\theta_n$ for every $j\in\{1,...,d_g\}$. 
Therefore, $\lambda\in\sqrt{n}(\Theta-\theta_n)$ eventually. 

Conversely, let $\lambda\notin\Lambda$. 
Then, there exists a $j\in\{1,...,d_g\}$, such that $e'_jb+e'_jG(\theta_0)\lambda>0$. 
By continuity we can find an $\epsilon>0$ such that for all $\tilde b\in B(b,\epsilon)$, for all $\tilde\theta\in B(\theta_0,\epsilon)$, and for all $\tilde\lambda\in B(\lambda,\epsilon)$, $e'_j\tilde b+e'_jG(\tilde\theta)\tilde\lambda>0$. 
Next, note that we can define $\tilde\theta_n=\theta_n+n^{-1/2}\tilde\lambda$ for all $\tilde\lambda\in B(\lambda,\epsilon)$. 
It follows that $\tilde\theta_n\in B(\theta_0,\epsilon)$ eventually along any subsequence in $n$, where the eventuality is uniform over $\tilde\lambda\in B(\lambda,\epsilon)$. 
Therefore, by a mean value expansion, $\sqrt{n}e'_jg(\tilde\theta_n)=\sqrt{n}e'_jg(\theta_n)+e'_jG(\ddot\theta_n)\tilde\lambda>0$ for some $\ddot\theta_n$ between $\tilde\theta_n$ and $\theta_n$ eventually along any subsequence in $n$, where the eventuality is uniform over $\tilde\lambda\in B(\lambda,\epsilon)$. 
This implies that $\lambda\notin \limsup_{n\rightarrow\infty}\sqrt{n}(\Theta-\theta_n)$. 

The above arguments imply 
\begin{equation}
\Lambda^o\subset \liminf_{n\rightarrow\infty} \sqrt{n}(\Theta-\theta_n)\subset \limsup_{n\rightarrow\infty}\sqrt{n}(\Theta-\theta_n)\subset \Lambda. 
\end{equation}
The lemma then follows from taking the closure, using the fact that the PK liminf and limsup are always closed. 
\end{proof}

\subsection{Proofs for Section \ref{Weak_Identification_Section}}

\begin{proof}[Proof of Lemma \ref{weakid-K-convergence}]
\textup{\textbf{(a)}} Let $\lambda=(\beta,\lambda_\pi)\in\{\beta\in\R^{d_\beta}:(\beta,\pi_0)\in int(\Theta)\}\times\R^{d_\pi}$. 
Then there exists an $\epsilon>0$ so that $B(\beta,\epsilon)\times B(\pi_0,\epsilon)\subset\Theta$. 
Eventually, $\pi_n+n^{-1/2}\lambda_\pi\in B(\pi_0,\epsilon)$. 
Therefore $\lambda\in\Lambda^{\hspace{-0.48mm}\text{W}}_n$ eventually. 

Next, let $\lambda=(\beta,\lambda_\pi)\notin \mathcal{B}^{\text{W}}\times \R^{d_\pi}$. 
Since $\Theta$ is closed, there exists an $\epsilon>0$ such that $B(\beta,\epsilon)\times B(\pi_0,\epsilon)\cap \Theta=\emptyset$. 
For any $\tilde\lambda_\pi\in B(\lambda_\pi,\epsilon)$, $\pi_n+n^{-1/2}\tilde\lambda_\pi\in B(\pi_0,\epsilon)$ eventually along any subsequence in $n$, where the eventuality is uniform over $\tilde\lambda_\pi\in B(\lambda_\pi,\epsilon)$. 
This implies that $\lambda\notin\limsup_{n\rightarrow\infty}\Lambda_n^{\hspace{-0.48mm}\text{W}}$. 

The above arguments imply
\begin{equation}
\{\beta\in\R^{d_\beta}: (\beta,\pi_0)\in int(\Theta)\}\times\R^{d_\pi}\subset \liminf_{n\rightarrow\infty}\Lambda_n^{\hspace{-0.48mm}\text{W}}\subset\limsup_{n\rightarrow\infty}\Lambda_n^{\hspace{-0.48mm}\text{W}}\subset\mathcal{B}^{\text{W}}\times\R^{d_\pi}. 
\end{equation}
The result then follows by taking the closure, using the fact that $\mathcal{B}^{\text{W}}=cl(\{\beta\in\R^{d_\beta}:(\beta,\pi_0)\in int(\Theta)\})$ and that the PK liminf and limsup are always closed. 

\textup{\textbf{(b)}} The condition on $\tilde\lambda$ implies the interior of $\mathcal{B}^{\text{SS}}$ is given by $\mathcal{B}^o=\{\tilde\lambda\in\R^{d_\beta}: e'_jb+e'_j G_\beta(\beta_0,\pi_0)\tilde\lambda<0 \text{ for all }j\in\{1,...,d_g\}\}$. 
It also implies that $\mathcal{B}^{\text{SS}}$ is equal to the closure of its interior. 

Let $\lambda=(\lambda_\beta,\lambda_\pi)\in \mathcal{B}^o\times \R^{d_\pi}$. 
Note that by a mean value expansion, $\tilde\theta_n=(\beta_n+a_n^{-1}\lambda_\beta,\pi_n+n^{-1/2}\lambda_\pi)$ satisfies
\begin{align}
a_n e'_j g(\tilde\theta_n)= & a_n e'_jg(\beta_n,\pi_n)+e'_j G_\beta(\ddot{\theta}_n)\lambda_\beta+a_n n^{-1/2}e'_j G_\pi(\ddot{\theta}_n)\lambda_\pi\\
\rightarrow & e'_j b+e'_j G_\beta(\beta_0,\pi_0)\lambda_\beta<0 \nonumber
\end{align}
for some $\ddot\theta_n$ between $\tilde\theta_n$ and $\theta_n$ for every $j\in\{1,...,d_g\}$. 
Therefore, $\lambda\in \Lambda_n^{\text{SS}}$ eventually. 

Conversely, let $\lambda=(\lambda_\beta,\lambda_\pi)\notin\mathcal{B}^{\text{SS}}\times \R^{d_\pi}$. 
Then, there exists a $j\in\{1,...,d_g\}$, 
such that $e'_jb+e'_jG_\beta(\beta_0,\pi_0)\lambda_\beta>0$. 
By continuity we can find an $\epsilon>0$ such that for all $\tilde b\in B(b,\epsilon)$, for all $\tilde\theta\in B(\theta_0,\epsilon)$, and for all $\tilde\lambda_\beta\in B(\lambda_\beta,\epsilon)$, $e'_j\tilde b+e'_jG_\beta(\tilde\theta)\tilde\lambda_\beta>0$. 
Next note that we can define $\tilde\theta_n=(\beta_n+a_n^{-1}\tilde\lambda_\beta,\pi_n+n^{-1/2}\tilde\lambda_\pi)$ for all $\tilde\lambda_\beta\in B(\lambda_\beta,\epsilon)$ and $\tilde\lambda_\pi\in B(\lambda_\pi,\epsilon)$. 
It follows that $\tilde\theta_n\in B(\theta_0,\epsilon)$ eventually along any subsequence in $n$, where the eventuality is uniform over $(\tilde\lambda_\beta,\tilde\lambda_\pi)\in B(\lambda_\beta,\epsilon)\times B(\lambda_\pi,\epsilon)$. 
Therefore, by a mean value expansion, $a_ne'_jg(\tilde\theta_n)=a_ne'_jg(\beta_n,\pi_n)+e'_jG_\beta(\ddot\theta_n)\tilde\lambda_\beta+a_nn^{-1/2}e'_jG_\pi(\ddot\theta_n)\tilde\lambda_\pi>0$ for some $\ddot\theta_n$ between $\tilde\theta_n$ and $\theta_n$ eventually along any subsequence in $n$, where the eventuality is uniform over $(\tilde\lambda_\beta,\tilde\lambda_\pi)\in B(\lambda_\beta,\epsilon)\times B(\lambda_\pi,\epsilon)$. 
This implies that $\lambda\notin \limsup_{n\rightarrow\infty}\Lambda_n^{\text{SS}}$. 

The above arguments imply 
\begin{equation}
\mathcal{B}^o\times \R^{d_\pi}\subset\liminf_{n\rightarrow\infty}\Lambda_n^{\text{SS}}\subset\limsup_{n\rightarrow\infty}\Lambda_n^{\text{SS}}\subset \mathcal{B}^{\text{SS}}\times \R^{d_\pi}. 
\end{equation}
The result then follows by taking the closure, using the fact that the PK liminf and limsup are always closed. 
\end{proof}

\subsection{Proofs and Lemmas for Section \ref{Proof_of_Thm_1_Section}}

The proofs of Lemmas \ref{J.14_version_2} and \ref{J.12} rely on additional lemmas stated and proved at the end of this section. 

\begin{proof}[Proof of Lemma \ref{J.14_version_2}]
We first prove the existence of $K_1$. 
Note that $\Lambda_n\cap K$ is a sequence of subsets of $K$, which is a compact and therefore separable metric space. 
By Theorem 1.1.7 in \cite{AubinFrankowska1990}, there exists a subsequence, $n_q$, and a compact set $K_1\subset K$, such that $\Lambda_{n_q}\cap K\rightarrow K_1$. 
We next confirm that $K_1\subset \Lambda$. 
For any $h\in K_1$, the fact that $\Lambda_{n_q}\cap K\rightarrow K_1$ implies that there exists a sequence $h_q\in\Lambda_{n_q}\cap K$ such that $h_q\rightarrow h$. 
The fact that $\Lambda_{n_q}\rightarrow \Lambda$ implies that $h\in\Lambda$. 
Therefore, $K_1\subset \Lambda$, and we can write $K_1=\Lambda\cap K_1$. 

We next prove the existence of $K_3$. 
For simplicity, re-index the above subsequence by $n$ instead of $n_q$. 
By Lemma \ref{KuratowskiEquivalence2}, $\vec d(\Lambda\cap K,\Lambda_n)\rightarrow 0$. 
Define an increasing sequence of positive integers, $N_m$, such that for every $m\in\N$, $\vec d(\Lambda\cap K,\Lambda_n)<m\inv$ for all $n\ge N_m$. 
For each $m\in\N$, let $U_m=\cup_{h\in\Lambda\cap K}B(h,m^{-1})$. 
Next, for each $m\in\N$ and for each $n\in\{N_m, N_m+1, ..., N_{m+1}-1\}$, we note that $\{B(h,m\inv): h\in\Lambda_n\cap U_m\}$ is an open cover of $\Lambda\cap K$. 
Let $F_n\subset\Lambda_n$ denote a finite set composed of the centers of the balls that index a finite subcover. 
Let $F=\cup_{n=N_1}^\infty F_n$ and let $K_3=F\cup K$. 

We show that $K_3$ is compact. 
Let $h_q$ be a sequence in $K_3$. 
If there exists a subsequence that belongs to $K$, then there exists a further subsequence that converges to a point in $K$. 
Thus, we consider the case that $h_q\in F$ eventually. 
If there exists an $N$ such that $h_q$ is in $\cup_{n=N_1}^N F_n$ along a subsequence, then by finiteness, there exists a subsequence that is constant, and thus converges. 
Thus, it is sufficient to consider a sequence such that $h_q\in F_{n_q}$, where $n_q\rightarrow\infty$. 
We can take a subsequence, $h_m=h_{q_m}$, so that for all $m$, $n_{q_m}\ge N_m$. 
Then, for every $m$, there exists $\tilde h_m\in K$ such that $d(\tilde h_m,h_m)<m\inv$ because $F_{n_{q_m}}\subset U_m=\cup_{h\in\Lambda\cap K} B(h,m^{-1})$. 
Because $K$ is compact, $\tilde h_m$ has a limit point in $K$, and $h_m$ must have the same limit point. 
Therefore, $h_m$ converges along a subsequence to a point in $K$. 
We have shown that every sequence in $K_3$ has a subsequence that converges to a point in $K_3$. 
Therefore, $K_3$ is compact. 

We repeat the argument in the first paragraph to get a subsequence, $n_q$, and a compact set $\widetilde K_2$ such that $\Lambda_{n_q}\cap K_3\rightarrow \widetilde K_2$ and $\widetilde K_2\subset\Lambda\cap K_3$. 
To complete the proof, we show that $\Lambda\cap K\subset \widetilde K_2$. 
If this is true, then the lemma follows from taking $K_2=\widetilde K_2\cup K$ because then $\widetilde K_2=\Lambda\cap K_2$. 
Let $h\in\Lambda\cap K$ and let $\epsilon>0$. 
Let $m\in\N$ such that $m\inv<\epsilon$. 
For every $n\ge N_m$, we have $\Lambda\cap K\subset \cup_{h\in F_n}B(h,m\inv)$, so there exists a $\tilde h\in F_n\subset \Lambda_n\cap K_3$ such that $d(h,\tilde h)<m\inv<\epsilon$. 
This implies that there exists a subsequence, $n_q$, and a sequence, $\tilde h_q\in \Lambda_{n_q}\cap K_3$ such that $\tilde h_q\rightarrow h$. 
We can take this subsequence to be a further subsequence of the one for which $\Lambda_{n_q}\cap K_3\rightarrow \widetilde K_2$. 
Therefore, $h\in\widetilde K_2$. 
\end{proof}

\begin{proof}[Proof of Lemma \ref{J.12}]
We prove Lemma \ref{J.12} under the conditions of the furthermore. 
The main part of the lemma follows from the furthermore with $\mathcal{D}$ trivial. 

Let $\rho$ denote the extended Prokhorov metric on probability distributions over $\mathbb{R}\times\mathcal{D}$, as defined in Section 3.6 in \cite{Dudley2014}. 
Note that $\rho$ metrizes weak convergence by Theorem 3.28 in \cite{Dudley2014}. 

Let $\epsilon>0$ and let $n_m$ denote an arbitrary subsequence. 
Let $K$ be a compact set such that $P(\hat h \in K)> 1-\epsilon$ and $P_*(\hat h_n\in K)> 1-\epsilon$ eventually, where $P_\ast$ denotes the inner probability measure. 
By Lemma \ref{J.14_version_2}, there exist compact sets, $K\subset K_2\subset K_3$, and a further subsequence $n_q$ such that $\Lambda_{n_q}\cap K_3\rightarrow \Lambda\cap K_2$. 
We use the extended continuous mapping theorem to show that $\sup_{h\in\Lambda_{n_q}\cap K_3}\M_{n_q}(h)\rightsquigarrow \sup_{h\in \Lambda\cap K_2}\M(h)$. 
The condition of the extended continuous mapping theorem is satisfied by Lemma \ref{ECMT}. 
Under the conditions of the furthermore, this holds jointly with $\mathbb{D}_{n_q}\rightsquigarrow\mathbb{D}$. 
Therefore, 
\begin{equation}
\rho\left(\left(\sup_{h\in\Lambda_{n_q}\cap K_3}\M_{n_q}(h),\mathbb{D}_{n_q}\right), \left(\sup_{h\in\Lambda\cap K_2}\M(h),\mathbb{D}\right)\right)<\epsilon \label{J.12c}
\end{equation}
eventually. 

Let $F$ denote a closed nonempty set in $\R\times \mathcal{D}$ and let $F^\epsilon$ denote an epsilon expansion: $\{(x,D)\in\R\times \mathcal{D}: d((x,D),F)<\epsilon\}$. 
Calculate 
\begin{align}
&P\left(\left(\sup_{h\in\Lambda}\M(h),\mathbb{D}\right)\in F\right)-P_*\left(\left(\sup_{h\in\Lambda_{n_q}}\M_{n_q}(h),\mathbb{D}_{n_q}\right)\in F^{3\epsilon}\right)\nonumber\\
=&P\left(\left(\sup_{h\in\Lambda}\M(h),\mathbb{D}\right)\in F\right)-P\left(\left(\sup_{h\in\Lambda\cap K_2}\M(h),\mathbb{D}\right)\in F^\epsilon\right)\label{Prohorov1}\\
&+ P\left(\left(\sup_{h\in\Lambda\cap K_2}\M(h),\mathbb{D}\right)\in F^\epsilon\right)-P_*\left(\left(\sup_{h\in\Lambda_{n_q}\cap K_3}\M_{n_q}(h),\mathbb{D}_{n_q}\right)\in F^{2\epsilon}\right)\label{Prohorov2}\\
&+P_*\left(\left(\sup_{h\in\Lambda_{n_q}\cap K_3}\M_{n_q}(h),\mathbb{D}_{n_q}\right)\in F^{2\epsilon}\right)-P_*\left(\left(\sup_{h\in\Lambda_{n_q}}\M_{n_q}(h),\mathbb{D}_{n_q}\right)\in F^{3\epsilon}\right).\label{Prohorov3}
\end{align}
We show that each of lines (\ref{Prohorov1}) to (\ref{Prohorov3}) is small. 
Note that (\ref{Prohorov2}) is less than $\epsilon$ eventually by (\ref{J.12c}). 

Line (\ref{Prohorov1}) satisfies 
\begin{align}
&P\left(\left(\sup_{h\in\Lambda}\M(h),\mathbb{D}\right)\in F\right)-P\left(\left(\sup_{h\in\Lambda\cap K_2}\M(h),\mathbb{D}\right)\in F^\epsilon\right)\\
\le&P\left(\{\left(\sup_{h\in\Lambda}\M(h),\mathbb{D}\right)\in F\}\cap\{\left(\sup_{h\in\Lambda\cap K_2}\M(h),\mathbb{D}\right)\notin F^\epsilon\}\right)\nonumber\\
\le&P\left(\sup_{h\in\Lambda\cap K_2}\M(h)<\sup_{h\in\Lambda}\M(h)-\epsilon\right)\nonumber\\
\le&P\left(\hat h\notin K_2\right)\le P\left(\hat h\notin K\right)<\epsilon, \nonumber
\end{align}
where the first inequality follows from $P(A)-P(B)\le P(A\cap B^c)$, the second inequality follows from the $\epsilon$ expansion of $F$, the third inequality follows because $\hat h$ achieves the maximum over $h\in\Lambda$, the fourth inequality follows because $K\subset K_2$, and the final equality follows from the choice of $K$. 

Line (\ref{Prohorov3}) satisfies 
\begin{align}
&P_*\left(\left(\sup_{h \in \Lambda_{n_q}\cap K_3}\M_{n_q}(h),\mathbb{D}_{n_q}\right)\in F^{2\epsilon}\right)-P_*\left(\left(\sup_{h\in\Lambda_{n_q}}\M_{n_q}(h),\mathbb{D}_{n_q}\right)\in F^{3\epsilon}\right)\\
\le&P^*\left(\{\left(\sup_{h\in\Lambda_{n_q}\cap K_3}\M_{n_q}(h),\mathbb{D}_{n_q}\right)\in F^{2\epsilon}\}\cap\{\left(\sup_{h\in\Lambda_{n_q}}\M_{n_q}(h),\mathbb{D}_{n_q}\right)\notin F^{3\epsilon}\}\right)\nonumber\\
\le&P^*\left(\sup_{h\in\Lambda_{n_q}\cap K_3}\M_{n_q}(h)<\sup_{h\in\Lambda_{n_q}}\M_{n_q}(h)-\epsilon\right)\nonumber\\
\le&P^*\left(\hat h_{n_q}\notin K_3\right)+P^*(o_P(1)\ge\epsilon)<2\epsilon\nonumber
\end{align}
eventually, where the first inequality follows from Lemma \ref{OuterMeasure}, the second inequality follows from the $\epsilon$ expansion of $F^{2\epsilon}$, the third inequality follows from the assumed condition on $\hat h_n$, and the final inequality holds eventually by the choice of $K$, using $K\subset K_3$. 

Lines (\ref{Prohorov1}) to (\ref{Prohorov3}) imply 
\begin{equation}
\rho\left(\left(\sup_{h\in\Lambda_{n_q}}\M_{n_q}(h),\mathbb{D}_{n_q}\right), \left(\sup_{h\in\Lambda}\M(h),\mathbb{D}\right)\right)<4\epsilon \label{J.12d}
\end{equation}
eventually. 
We have shown that for every subsequence, $n_m$, there exists a further subsequence, $n_q$, such that (\ref{J.12d}) holds eventually. 
Therefore, (\ref{J.12d}) holds along the original sequence eventually. 
The fact that $\epsilon>0$ was arbitrary implies convergence in the Prokhorov metric, which implies weak convergence by Theorem 3.28 in \cite{Dudley2014}. 
\end{proof}

\begin{lemma}
\label{KuratowskiEquivalence2}
Let $H$ be a metric space equipped with metric $d$. 
Suppose $\Lambda_n$ and $\Lambda$ are subsets of $H$  such that $\Lambda_n\rightarrow\Lambda$. 
Then, for any compact set, $K$, $\vec d(\Lambda\cap K,\Lambda_n)\rightarrow 0$ and $\vec d(\Lambda_n\cap K,\Lambda)\rightarrow 0$. 
\end{lemma}
\begin{proof}[Proof of Lemma \ref{KuratowskiEquivalence2}]
Recall $\vec d(A,B)=\max(\sup_{a\in A}\inf_{b\in B}d(a,b),0)$. 
We first show that $\vec d(\Lambda\cap K,\Lambda_n)\rightarrow 0$. 
Let $h_{1n}\in \Lambda\cap K$ denote a sequence that approximates the sup in $\vec d(\Lambda\cap K,\Lambda_n)$. 
(If $\Lambda\cap K$ is empty, then $\vec d(\Lambda\cap K,\Lambda_n)=0$ for all $n$.) 
By compactness, there exists a subsequence, $n_q$, such that $h_{1n_q}\rightarrow h_1\in \Lambda\cap K$. 
Since $h_1\in\Lambda$, by the definition of PK convergence, there exists a sequence, $h_{2n}\in\Lambda_n$ such that $h_{2n}\rightarrow h_1$. 
This means that $\vec d(\Lambda\cap K,\Lambda_n)=d(h_{1n},\Lambda_n)+o(1)\le d(h_{1n},h_{2n})+o(1)\rightarrow 0$. 

We next show that $\vec d(\Lambda_n\cap K,\Lambda)\rightarrow 0$. 
Let $n_m$ be an arbitrary subsequence. 
Let $h_{1n_m}\in \Lambda_{n_m}\cap K$ denote a sequence that approximates the sup in $\vec d(\Lambda_{n_m}\cap K,\Lambda)$. 
(If $\Lambda_{n_m}\cap K$ is eventually empty, then $\vec d(\Lambda_{n_m}\cap K,\Lambda)=0$. Otherwise, take a further subsequence, $n_p$, so that $\Lambda_{n_p}\cap K$ is nonempty for all $p$.) 
By compactness, there exists a further subsequence, $n_q$, such that $h_{1n_q}\rightarrow h_1\in K$. 
By the definition of PK convergence, $h_1\in\Lambda$. 
Therefore, $\vec d(\Lambda_{n_q}\cap K,\Lambda)=d(h_{1n_q},\Lambda)+o(1)\le d(h_{1n_q},h_1)+o(1)\rightarrow 0$. 
This shows that for every subsequence, $n_m$, there exists a further subsequence, $n_q$, such that $\vec d(\Lambda_{n_q}\cap K,\Lambda)\rightarrow 0$. 
This implies that $\vec d(\Lambda_n\cap K,\Lambda)\rightarrow 0$. 
\end{proof}

\begin{lemma}
\label{OuterMeasure}
Let $(\Omega,\mathcal{F},P)$ be a probability space. 
Let $P^*$ denote the outer measure, and let $P_*$ denote the inner measure. 
Let $A$ and $B$ be two (possibly non-measurable) subsets of $\Omega$. 
Then $P_*(A\cup B)\le P_*(A)+P^*(B)$. 
\end{lemma}
\begin{proof}[Proof of Lemma \ref{OuterMeasure}]
Let $M_A$ denote a measurable set contained in $A$, let $M_B$ denote a measurable set containing $B$, and let $M_\cup$ denote a measurable set contained in $A\cup B$. 
For any fixed $M_B$ and $M_\cup$, note that 
\begin{equation}
\sup_{M_A\subset A} P(M_A)+P(M_B)\ge P(M_{\cup}/M_B)+P(M_B)\ge P(M_{\cup}), 
\end{equation}
where $M_{\cup}/M_B=\{x\in M_{\cup}: x\notin M_B\}$. 
The result follows from taking the sup over $M_{\cup}\subset A\cup B$ and the inf over $M_B\supset B$. 
\end{proof}

\begin{lemma}
\label{ECMT}
Let $K$ be a compact metric space. 
Let $\Lambda_n$ be a sequence of subsets of $K$ that PK converges to $\Lambda\subset K$. 
Let $g_n\in \ell^\infty(K)$ be a sequence of functions on $K$ converging uniformly to a continuous function, $g\in\ell^\infty(K)$. 
Then, 
\[
\left|\sup_{x\in\Lambda_n}g_n(x)-\sup_{x\in\Lambda}g(x)\right|\rightarrow 0. 
\]
\end{lemma}
\begin{proof}[Proof of Lemma \ref{ECMT}]
First, if $\Lambda$ is empty, then $\sup_{x\in\Lambda_n}g_n(x)=\sup_{x\in\Lambda}g(x)=-\infty$ eventually because $\Lambda_n$ is empty eventually. 
Otherwise, if $\Lambda$ is nonempty, then $\Lambda_n$ is nonempty eventually. 
In that case, notice that 
\begin{align}
\left|\sup_{x\in\Lambda_n}g_n(x)-\sup_{x\in\Lambda}g(x)\right|\le&\left|\sup_{x\in\Lambda_n}g_n(x)-\sup_{x\in\Lambda_n}g(x)\right|+\left|\sup_{x\in\Lambda_n}g(x)-\sup_{x\in\Lambda}g(x)\right|\\
\le&\sup_{x\in K} \left|g_n(x)-g(x)\right|+\left|\sup_{x\in\Lambda_n}g(x)-\sup_{x\in\Lambda}g(x)\right|, \nonumber
\end{align}
where the first inequality follows by the triangle inequality, and the second inequality follows from Lemma \ref{uniformbound}. 
The first term goes to zero because $g_n(x)\rightarrow g(x)$ uniformly over $K$. 
The second term goes to zero because $g(x)$ is uniformly continuous over $K$. 
That is, for any $\epsilon>0$, let $\delta>0$ so that $d(x,y)<\delta$ implies $|g(x)-g(y)|<\epsilon$. 
By Lemma \ref{KuratowskiEquivalence2}, $\vec d(\Lambda,\Lambda_n)<\delta$ and $\vec d(\Lambda_n,\Lambda)<\delta$ eventually. 
Then, for any $x\in\Lambda$, $g(x)\le \sup_{x\in\Lambda_n}g(x)+\epsilon$. 
Taking the sup gives $\sup_{x\in\Lambda}g(x)\le \sup_{x\in\Lambda_n}g(x)+\epsilon$. 
A similar argument gives $\sup_{x\in\Lambda_n}g(x)\le \sup_{x\in\Lambda}g(x)+\epsilon$. 
Thus, the second term converges to zero. 
\end{proof}

\begin{lemma}
\label{uniformbound}
If $f$ and $g$ are real-valued bounded functions on a nonempty set $X$, then 
\[
|\sup_{x\in X}f(x)-\sup_{x\in X}g(x)|\le \sup_{x\in X}|f(x)-g(x)|. 
\]
\end{lemma}
\begin{proof}[Proof of Lemma \ref{uniformbound}]
Let $\epsilon>0$. 
Suppose $\sup_{x\in X}f(x)\ge \sup_{x\in X}g(x)$ without loss of generality. 
Let $\bar x\in X$ such that $f(\bar x)\ge \sup_{x\in X}f(x)-\epsilon$. 
Then 
\begin{align}
|\sup_{x\in X}f(x)-\sup_{x\in X}g(x)|=&\sup_{x\in X}f(x)-\sup_{x\in X}g(x)\\
\le&f(\bar x)-\sup_{x\in X}g(x)+\epsilon\nonumber\\
\le&f(\bar x)-g(\bar x)+\epsilon\nonumber\\
\le&\sup_{x\in X}|f(x)-g(x)|+\epsilon. \nonumber
\end{align}
Taking $\epsilon$ to zero proves the result. 
\end{proof}

\section*{Funding Source}

This research did not receive any specific grant from funding agencies in the public, commercial, or not-for-profit sectors. 

\section*{Acknowledgments}

This paper has benefited from helpful feedback from James Martin and Adam McCloskey. 

\bibliography{references}

\end{document}